\documentclass[12pt]{article}
\usepackage{ssi}
\usepackage{times}
\usepackage{epsfig}                   % please use epsfig for figures
\newif\ifhepex \hepextrue
\newif\ifnohepex \nohepextrue
\ifhepex \nohepexfalse \fi

\newcommand{\mydef}[2]{\def #1{\ifmmode{#2}
          \else ${#2}$\fi}}

\mydef{\degree}{^\circ}
\mydef{\BR}{{\cal B}}
\def\aDz{\ifmmode{\mathrm{\overline{D}}^{\: 0}}\else ${\mathrm{\overline{D}}^{\:
 0}}$\fi}
\def\Dz{\ifmmode{\mathrm{D}^{0}}\else ${\mathrm{D}^{0}}$\fi}
\def\gDz{\Dz}
\def\aDstz{\ifmmode{\mathrm{\overline{D}}^{\:\ast 0}}\else ${\mathrm{\overline{D
}}^{\:\ast 0}}$\fi}
\def\Dstz{\ifmmode{\mathrm{D}^{\ast 0}}\else ${\mathrm{D}^{\ast 0}}$\fi}
\def\Bz{\ifmmode{{\mathrm{B}^0}}\else ${\mathrm{B}^0}$\fi}
\def\aBz{\ifmmode{\mathrm{\overline{B}}^0}\else ${\mathrm{\overline{B}}^0}$\fi}
\def\gpz{\ifmmode{\mathrm{\pi^0}}\else ${\mathrm{\pi^0}}$\fi}
\mydef{\Ebeam}{E_{beam}}
\mydef{\yfours}{\Upsilon(4S)}
\mydef{\ythrees}{\Upsilon(3S)}
\mydef{\kreuz}{\times}
\mydef{\Jpsi}{J/\Psi}
\mydef{\JPsi}{J/\Psi}
\mydef{\jpsi}{\Jpsi}
\mydef{\upsones}{\Upsilon(1S)}
\mydef{\upstwos}{\Upsilon(2S)}
\mydef{\upsthrees}{\Upsilon(3S)}
\mydef{\upsfours}{\Upsilon(4S)}
\mydef{\yfours}{\Upsilon(4S)}
\mydef{\etab}{\eta_b}
\def\bar{\overline}

\newcommand{\up}[1]{\raisebox{1.5ex}[-1.5ex]{#1}}
\mydef{\ACP}{{\cal A}_{CP}}
\mydef{\Dtokpi}{\gDz\rightarrow K^- \pi^+}
\mydef{\Dtokpipi}{\gDz\rightarrow K^- \pi^+ \gpz}
\mydef{\Dtokpipipi}{\gDz\rightarrow K^- \pi^+\pi^+\pi^-}
\mydef{\Kstpl}{K^{*+}(892)}
\mydef{\mbc}{M_{B}}
\mydef{\mbcc}{\mbc}
\mydef{\deltae}{\Delta E}
\mydef{\FD}{{\cal F}_D}
\mydef{\ctst}{\cos\theta_{B-Hel.}}
\mydef{\Btodpi}{\aBz\rightarrow\Dz\gpz}
\mydef{\Btodstpi}{\aBz\rightarrow\Dstz\gpz}
\mydef{\BBar}{B\overline{B}}
\mydef{\Dsttodpi}{\Dstz\rightarrow\Dz\gpz}
\mydef{\Dsttodg}{\Dstz\rightarrow\Dz\gamma}
\mydef{\Navfit}{<N_{fitted}>}
\mydef{\Nexp}{N_{Expected}}
\mydef{\RMSfit}{RMS_{fit}}
\mydef{\Significance}{N/\sigma(N)}
\mydef{\Nsig}{N_{Sig}}
\mydef{\Nbb}{N_{B\overline{B}}}
\mydef{\Ncont}{N_{Cont}}
\mydef{\Like}{L}
\mydef{\DDstz}{D^{(*)0}}
\mydef{\gpp}{\pi^+}
\mydef{\gpz}{\pi^0}
\mydef{\gpm}{\pi^-}
\mydef{\gppm}{\pi^\pm}
\mydef{\lepton}{\ell}
\mydef{\Btoddstpi}{\aBz\rightarrow\DDstz\gpz}
\mydef{\to}{\rightarrow}
\mydef{\Btosg}{b\to s \gamma}
\mydef{\btosg}{b\to s \gamma}
\mydef{\Btokll}{B\to K^{(*)} {\lepton}^+{\lepton}^-}
\mydef{\kshort}{K_s^0}
\newcommand{\CKM}[1]{\ifmmode{\rm V_{#1}}\else ${\rm V_{#1}}$\fi}

\mydef{\lepton}{{\cal l}}
\mydef{\tilde}{\sim}
\newcommand{\etal}{\mbox{$et$ $al.$}}
\begin{document}

\title{Recent B Physics Results from CLEO}

\author{Eckhard von Toerne
\vskip 0.5in 
\noindent
Kansas State University,\\
Manhattan, KS 66506-2601\\
evt@phys.ksu.edu\\[0.4cm]
%and \\[0.4cm]
%Jane Eyre\thanks{Supported by NSF Contract xxx.}\\ 
%Washington University, Seattle, Washington xxxxx \\[0.4cm]
Representing the CLEO Collaboration
}

\maketitle
\begin{abstract}%
\baselineskip 16pt 
The CLEO detector is located at the
CESR ${\rm e^+e^-}$ collider in 
Ithaca, NY. CLEO's wide range of experimental measurements in b-hadron
decays is 
represented by improved measurements of $\CKM{cb}$ and
$\CKM{ub}$, rare B decays, and $b\bar{b}$ spectroscopy. 
New experimental results in
exclusive hadronic transitions will aid 
theorists in developing a theory of hadronic B decays. Such 
a theory will have 
consequences for the extraction of angles of the unitarity
triangle, especially $\gamma$. 
Recently, the CLEO collaboration 
has shifted its focus towards precision measurements at lower
energies. Based on
the new \ythrees\ data, we present 
the observation of a new bound $b\bar{b}$ state.
An outlook on the planned running at
${\rm\tau}$/charm-energies (CLEO-c) is given and the implications for
b-physics are discussed.
\ifhepex
\\[1cm]
\begin{center}
{\em Invited talk at\\ 
``Secrets of the B Meson'', SSI 2002 Topical Conference,\\
Stanford, CA, August 2002}
\end{center}
\fi
\end{abstract}
\section{Introduction}
\ifhepex
\pagestyle{plain}
\setcounter{page}{1}
\pagenumbering{arabic}
\fi
The advent of high-luminosity B factories and the discovery of
time-dependent CP asymmetries in the B-system \cite{babar_talk,belle_talk}
has transformed the whole field of B
physics. Precision tests of the standard model open up a window
for the discovery of new physics in B decays. This will require a
thorough understanding of time-dependent phenomena like mixing, and
also {\bf time-independent} phenomena like branching fractions and particle
spectra.   
Effects like final state interactions, 
re-scattering and interference between dominant
and suppressed decay amplitudes have to be understood. 
This makes it necessary to study extensively numerous rare and hadronic B
decays to gain full understanding of the dynamics\cite{heavy_flavor,browder_honscheid,factorization}. 

The CLEO collaboration has accumulated a large data set of 16
fb$^{-1}$ at the \yfours\ resonance with the CLEO-II, II.5 and III
detector configurations. 
Almost 50\% of data set were recorded 
with CLEO-III in a single year. This impressive
achievement proved to be insufficient to match the luminosity records of
the B-factories Babar and Belle.
CLEO returned to the $\Upsilon$ resonances below $B\bar{B}$ threshold 
(\upsones,\upstwos,\upsthrees) to collect data samples above or close to the 
total world data sets.
Most results presented here are based on the CLEO II and II.5 data. 
The integrated luminosity 
of this sub-sample is 9.1 fb$^{-1}$, collected on the \yfours\
resonance and 4.3 fb$^{-1}$ \tilde 60 MeV below the resonance to study the
continuum background from ${\rm e^+e^-\rightarrow q\bar{q}}$. The
importance of the 
{\em large} off-resonance sample lies in the background subtraction neccessary
in inclusive measurements such as $b\to s \gamma$ or the extraction of
$\CKM{ub}$ in the lepton energy endpoint region. 

\begin{table}[htb]
\begin{center}
\begin{tabular}{l|r|r|r}
\hline\hline
              & Resonance   & Continuum & $B \bar{B}$ \\
\up{Detector} & fb$^{-1}$   & fb$^{-1}$ & (10$^6$) \\
\hline
CLEO II       & 3.1         &   1.6     & 3.3 \\
CLEO II.V     & 6.0         &   2.8     & 6.4 \\
\hline        
Subtotal      & 9.1         &   4.4     & 9.7 \\
\hline
CLEO III (\yfours)     & 6.9         &  2.3      & 7.4 \\
\hline
Total (\yfours) & 16.0  & 6.7 & 17.1 \\
%\hline
%CLEO III (\ythrees)    & xxx & xxx & -- \\
\hline\hline
\end{tabular}
\end{center}
\parbox{145mm}{\caption[ ]{\label{nev_table}\it Integrated luminosities (on- and
off-resonance) and the number of $B \bar{B}$ pairs. }}
\end{table}
\begin{figure}[htb]
\ifhepex\vspace*{-0.8cm}\fi
\centerline{\epsfig{file=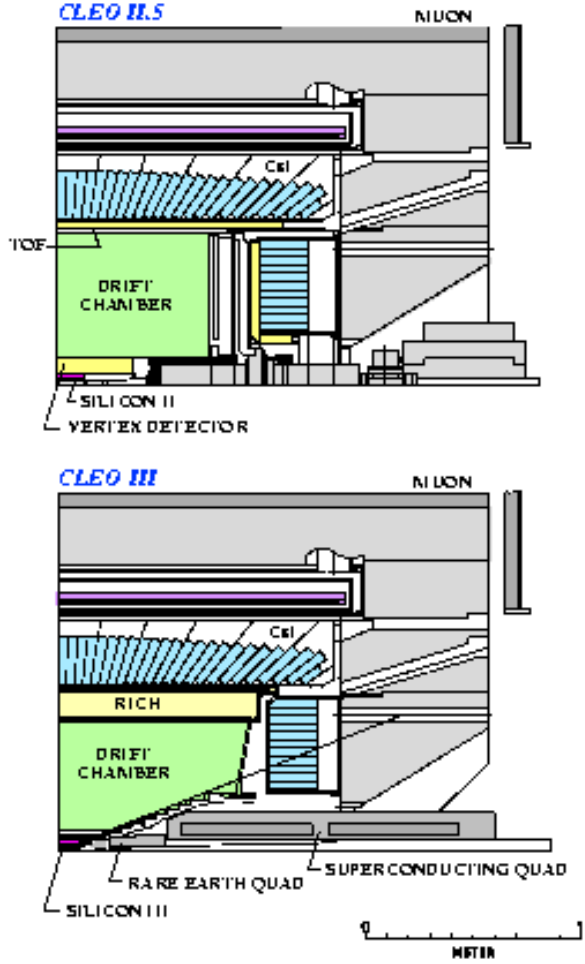,width=9.5cm}}
\begin{center}
\parbox{145mm}{\caption[ ]{\label{cleo_det.eps}\it  
Quarter sections of the CLEO-II.5 and CLEO-III detector configurations.}}
\end{center}
\vspace*{-0.5cm}
\end{figure}
%\clearpage

The CLEO II.5 and III detectors are shown
in Figure \ref{cleo_det.eps}.
The outer detector parts, the CsI calorimeter, superconducting coil, magnet iron and muon
chambers are common to all three detector configurations. In the CLEO
III upgrade, the CLEO II.5 silicon vertex detector, drift chamber and
time-of-flight counters were replaced by a new silicon vertex
detector, drift chamber, and a new Ring Imaging Cherenkov detector. 
Table \ref{nev_table} shows the integrated luminosities
obtained with each detector configuration. 

The kinematics of the \yfours\ decay, in which
two B mesons with equal masses are produced,
allow us to define two sensitive variables: 
the beam-constrained mass ${\rm \mbc = \sqrt{E^2_{\textstyle
beam} - P^2_{B}}}$ and the 
energy difference ${\rm \deltae = E_{B} - \Ebeam }$, where ${\rm
E_B}$ and ${\rm P_B}$ are the measured 
energy and momentum of the B candidate
and ${\rm E_{beam}}$ is the beam energy.

%There is a large number of recent CLEO publications 
%\cite{charm1}, \cite{charm2}, \cite{charm3} that
%are beyond the scope of this talk.

The CLEO collaboration and CESR plan to operate at center-of-mass
energies in the ${\rm\tau}$/charm region\cite{yellowbook}. This will
expand the scope of our on-going charm physics program and will allow
precision tests of perturbative QCD and lattice QCD predictions.
I will later explain the impact that these results, taken at lower
energy, will have on B physics.
%\clearpage
\section{Semi-Leptonic B decays}
\mydef{\Lam}{\bar{\Lambda}}
\mydef{\Lambar}{\bar{\Lambda}}
\mydef{\lamone}{\lambda_1}
\mydef{\lamtwo}{\lambda_2}
\mydef{\Tau}{{\cal T}}
\mydef{\Lamc}{\Lambda_c^+}
\mydef{\lamc}{\Lamc}
The partial semileptonic decay width 
${\rm \Gamma_{SL}^c = \Gamma(\overline{B}\to X_c\ell\overline{\nu})}$
is proportional to $|\CKM{cb}|^2$,  ${\rm \Gamma_{SL}^c = \gamma_c
|\CKM{cb}|^2}$, with the proportional factor $\gamma_c$ being dependent on
perturbative and non-perturbative parameters. 
The precision of the determination of $|\CKM{cb}|$ is mainly limited
by uncertainties on the parameters entering the expression
for $\gamma_c$.
Semi-leptonic rates and spectra can be expanded in a power series.
% with perturbative parameters \Lam\ \lamone\ and \lamtwo\  
To order $1/M_B^3$ the decay width is \cite{pdg_vcb}
\begin{eqnarray*} 
 \Gamma_{SL}^c = \frac{G_F^2 |\CKM{cb}|^2 M_B^5}{192 \pi^3}
(G_0+ 1/M_B G_1(\Lam) +  
1/M_B^2 G_2(\Lam, \lamone, \lamtwo) +\\ 
1/M_B^3 G_3(\Lam, \lamone, \lamtwo | {\scriptstyle \rho_1, \rho_2, \tau_1, \tau_2,
\tau_3, \tau_4})),
\end{eqnarray*}
with known functions $G_{0,1,2,3}$ and three main perturbative 
parameters $\Lam, \lamone, \lamtwo$, which are
accessible through experimental measurements. 
The parameter \lamone\ is related to the average kinetic energy of the
b-quark inside the B meson. 
The parameter \lamtwo\ is the expectation value of the leading
operator that breaks the heavy quark symmetry and can be determined
from the $B^*-B$ mass splitting. \Lam\ is related to the b-quark pole
mass $m_b$. The dependence on the remaining parameters $\rho_i,
\tau_j$ is expected to be relatively weak.

\begin{figure}[htb]
\centerline{\epsfig{file=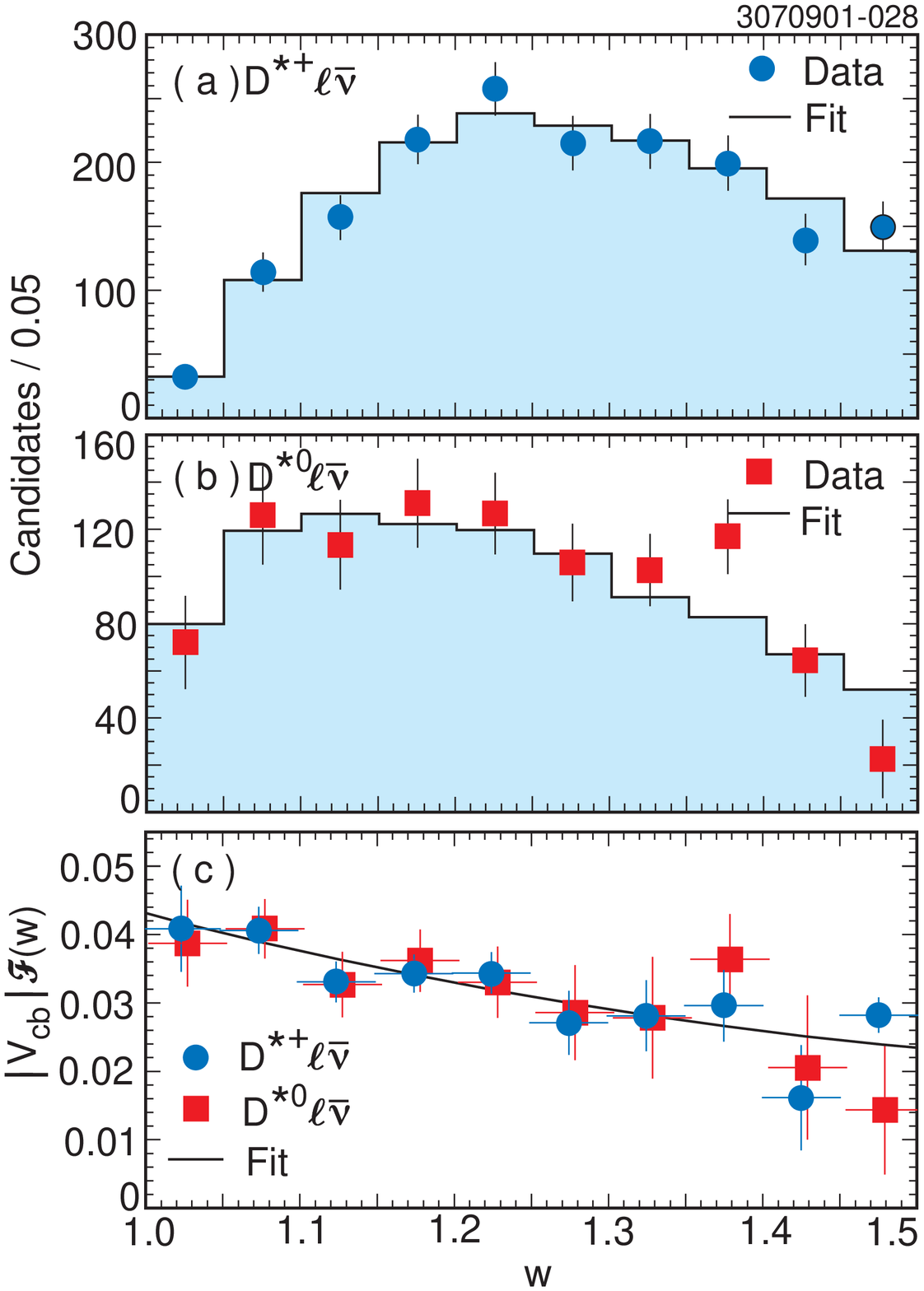,width=8cm}\epsfig{file=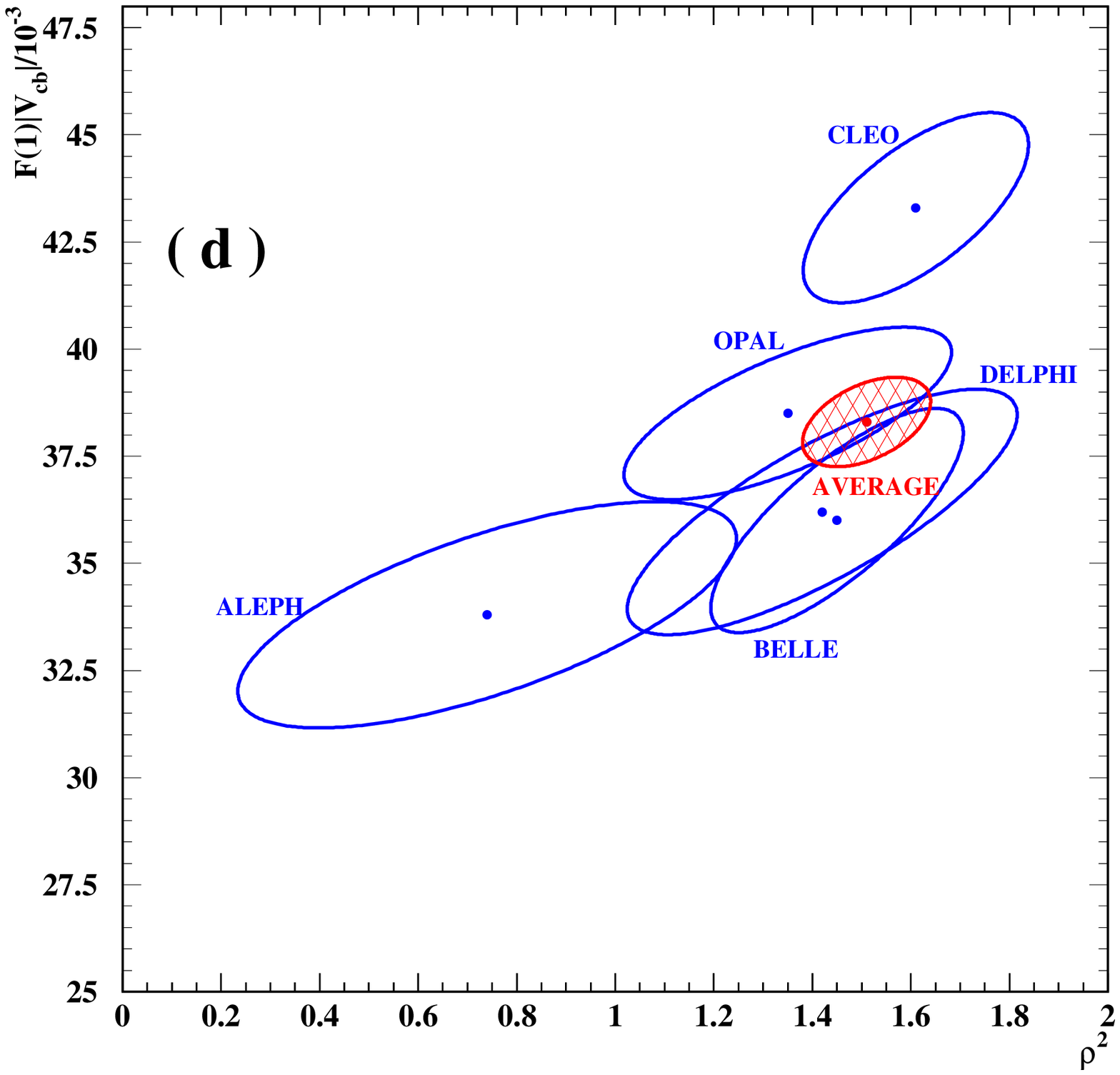,width=8cm}}
\begin{center}
\parbox{145mm}{\caption[ ]{\label{vcb_excl.eps}\it $\CKM{cb}$ in
exclusive decays. Signal yield
for $D^{*+}\ell\bar{\nu}$ (a) $D^{*0}\ell\bar{\nu}$ (b) unfolded
spectrum (c), comparison of fit results from different experiments (d).}}
\end{center}
\vspace*{-0.5cm}
\end{figure}

\paragraph{Moments of lepton spectra in semileptonic B decays.}
CLEO has pioneered measurements of moments of the hadronic mass spectrum in $B \to X_c
\ell \nu$ decays\cite{massmomentsold}. Our measurement together with
the measurement of moments in $b\to s \gamma$ \cite{bsg} allowed us to
extract the perturbative parameters \lamone\ and \Lambar.
\begin{figure}[htb]
\vspace*{-0.5cm}
\centerline{\epsfig{file=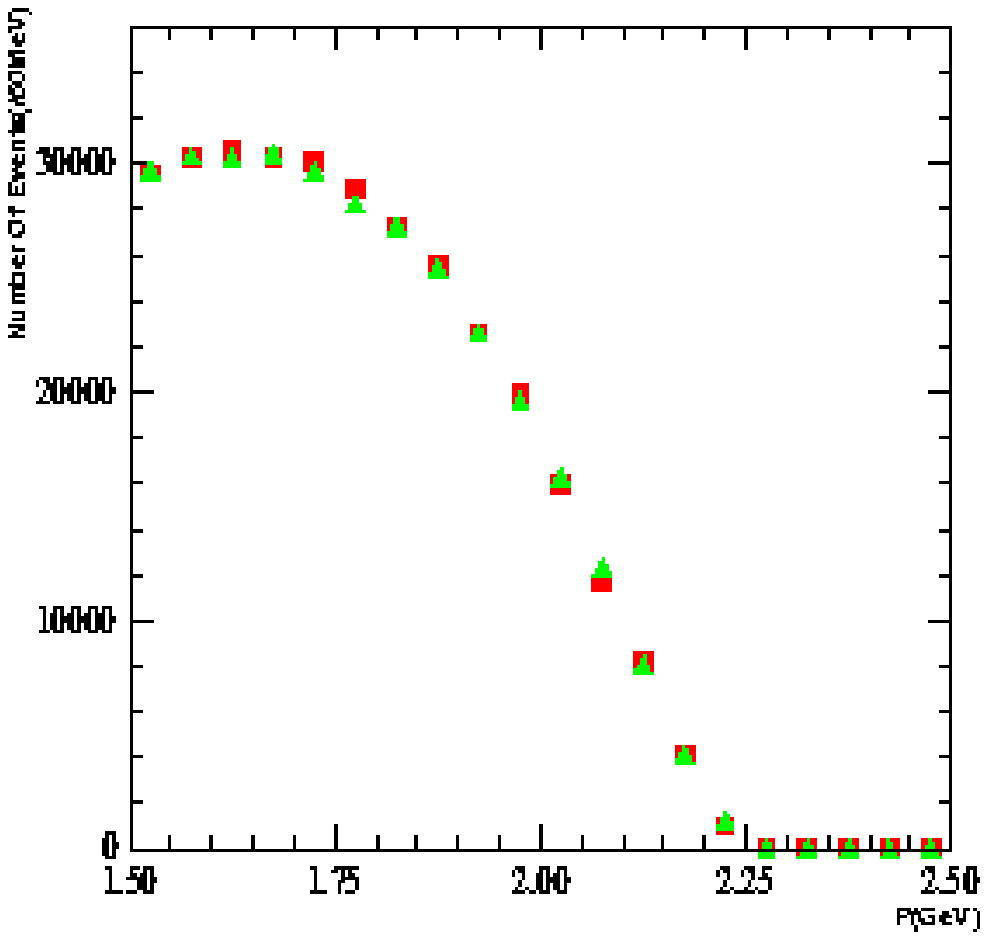,width=8cm}\epsfig{file=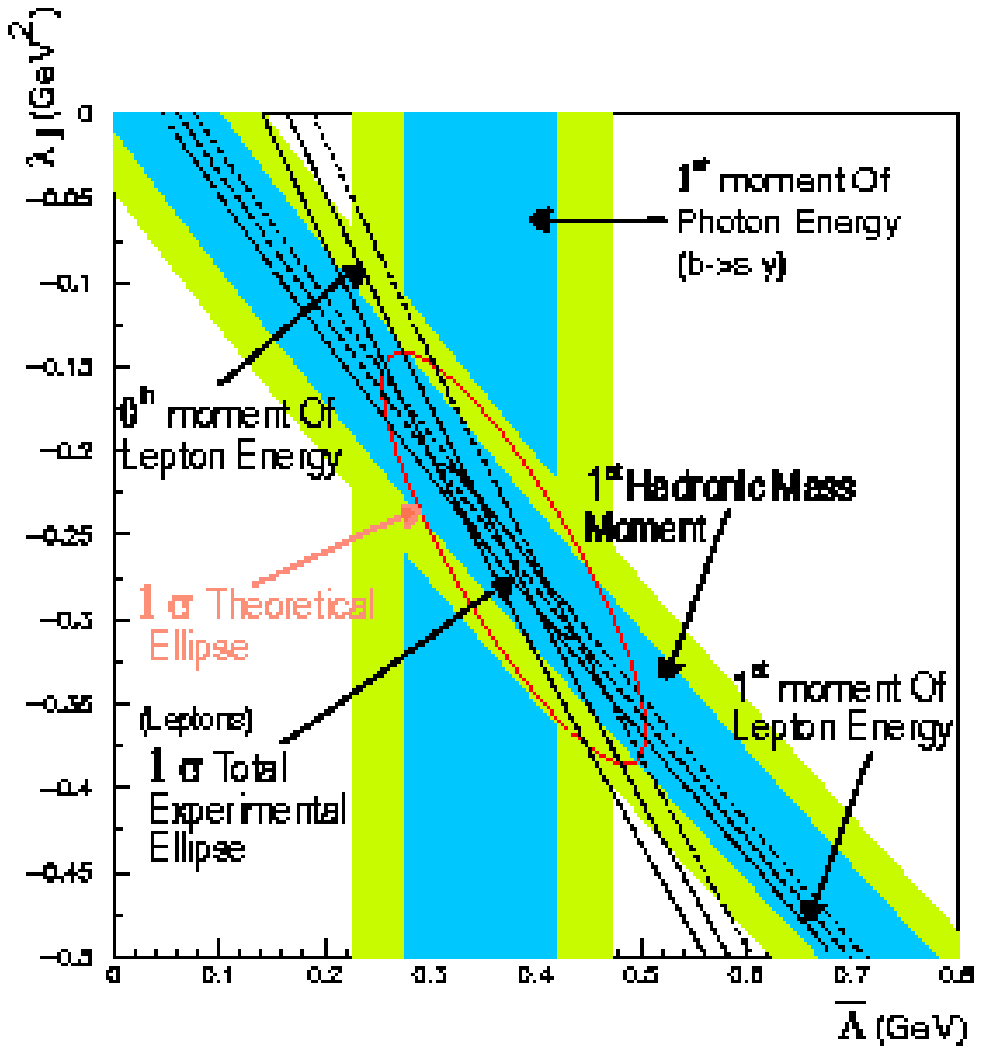,width=8cm}}
\begin{center}
\parbox{145mm}{\caption[ ]{\label{bothspct_atw.eps}\it {\bf (left)} Electron
(green triangles) and Muon spectra (red squares) above 1.5 GeV,
evaluated in the B-meson rest frame.
{\bf (right)} Constraints from the $\bar{B}\to X_c \ell\bar{\nu}$ hadronic
mass moments and $b\to s \gamma$ compared with the combined electron
and muon $R_0$ and $R_1$ results.}}
\end{center}
\vspace*{-0.5cm}
\end{figure}
A new CLEO result\cite{newmoments} involving
moments constitutes an
important cross check to our previous analysis. The lepton energy
spectrum has been analyzed following a suggestion from
M.~Gremm, A.~Kapustin, Z.~Ligeti and
M.~B.~Wise\cite{r1r2suggestion}. The two ratios extracted from the
data are 
\mydef{\gsl}{\Gamma_{sl}}
\mydef{\el}{E_\ell}
\[ R_0 =\frac{\int_{1.7}(d\gsl/d \el) d \el}{\int_{1.5}(d\gsl/d \el) d \el}\]
\[ R_1 =\frac{\int_{1.5}(\el d\gsl/d \el) d \el}{\int_{1.5}(d\gsl/d \el) d \el}\]
The lepton spectrum was truncated to
lepton momenta above 1.5 GeV in order to reduce the systematic
uncertainty due to secondary leptons from the cascade decays $b\to c
\to s$. The spectra for electrons and muons yield consistent results
(Fig.~\ref{bothspct_atw.eps}, left). 
The combined electron and muon result is
%measured ratios of truncated lepton 
\[ \Lambar=(0.39\pm0.03\pm0.06\pm0.12)\:GeV \: \:\: \lamone= (-0.25\pm0.02\pm0.05\pm0.14)\:GeV^2,\]
where the errors are statistical, systematic and theory error,
respectively.
The fact that the parameters extracted from lepton momentum spectra and
hadronic mass moments yield consistent results (as shown in
Fig.~\ref{bothspct_atw.eps}, right),  
represents a valuable cross check of the theory and its underlying assumptions. 
\paragraph{${\boldmath\CKM{cb}}$ from exclusive decays}
The decay $B \to D^* \ell \nu$ is a prime candidate for the extraction
of \CKM{cb} from exclusive decays. CLEO analyzes\cite{vcb_zerorecoil} 
$D^{*0}$ and $D^{*+}$ 
modes and obtains
$\BR(\aBz\to
D^{*+}\ell^-\bar{\nu}) = (6.09\pm 0.19\pm0.40)\times 10^{-2}$ and
$\BR(B^-\to D^{*0}\ell^-\bar{\nu}) = (6.50\pm 0.20\pm0.43)\times
10^{-2}$.
\mydef{\Wvcb}{{\cal W}}
We determine the yield as a function of $\Wvcb$, the boost of the $D^*$ in
the $B$ rest frame. The decay rate $d\Gamma/d\Wvcb$ extrapolated to
the kinematic endpoint ($\Wvcb=0$) can be calculated in Heavy Quark
Effective Theory and is proportional to  $|\CKM{cb}|^2$. 
The shape of $d\Gamma/d\Wvcb$ can be expressed with only one free
parameter $\rho$, which is approximately the slope of the
distribution in Fig.~\ref{vcb_excl.eps} (c). 
The two B decay modes give results that are 
consistent with each other.
The combined result is \[ |\CKM{cb}| = 0.0469 \pm
0.0014(stat)\pm0.0020(syst)\pm0.0018(theor.), \]
where the systematic error is dominated by the uncertainty on the
form factor calculation from lattice QCD, %\cite{vcb_formfactor}
${\cal{F}}(1)=0.92\pm0.03$. 

CLEO is the only experiment so far that has measured both the $D^{*+}$ and
the $D^{*0}$ decay modes. 
Our combined measurement is slightly higher than LEP and the B
factory measurements, employing the
same method for $D^{*+}$ only. 
The consistency of our result with these measurements
is at the 5\% level \cite{pdg_vcb} (Fig.~\ref{vcb_excl.eps}, d).
%\clearpage
\paragraph{${\boldmath\CKM{ub}}$ from inclusive decays}
The lepton endpoint region provides clear evidence for the existence
of $b\to u$ transitions. 
CLEO has published an updated measurement\cite{vub_inclusive} of $\CKM{ub}$ with 
the inclusive branching fraction. 
$\BR( B \to X_u \ell \nu)$, based on the CLEO II+II.5 data sets.
The measurement of $\BR( B \to X_u \ell \nu)$
depends on the successful removal of the dominating background due to
b $\to$ charm transitions. This can be achieved by exploiting the
larger kinematic range of $b\to u$ transitions, which restricts the
accessible $b\to u$ lepton spectrum to the endpoint region. 

The total uncertainty on $\CKM{ub}$ depends on the lepton
momentum range chosen. 
At low lepton momenta the huge background from $b\to
c$ transitions constitutes a large uncertainty. We chose the region of
$p_\ell=2.2-2.6\: GeV$ for our central value, which approximately
minimizes our total uncertainty.
We obtain an inclusive branching
fraction $\BR(B\to X_u \ell \nu)$ of
$(1.77\pm0.29\pm0.38)\times10^{-3}$, where the first error comes from
the branching fraction measurement and the second from the
extrapolation of the full momentum spectrum. This measurement
translates into a value 
\[ \CKM{ub} = (4.08 \pm
0.34\pm0.44\pm0.16\pm0.24)\times 10^{-3},\]
where the first two errors come from the branching fraction and the
third and fourth are theory contributions. 
 
\begin{figure}[htb]
\centerline{\epsfig{file=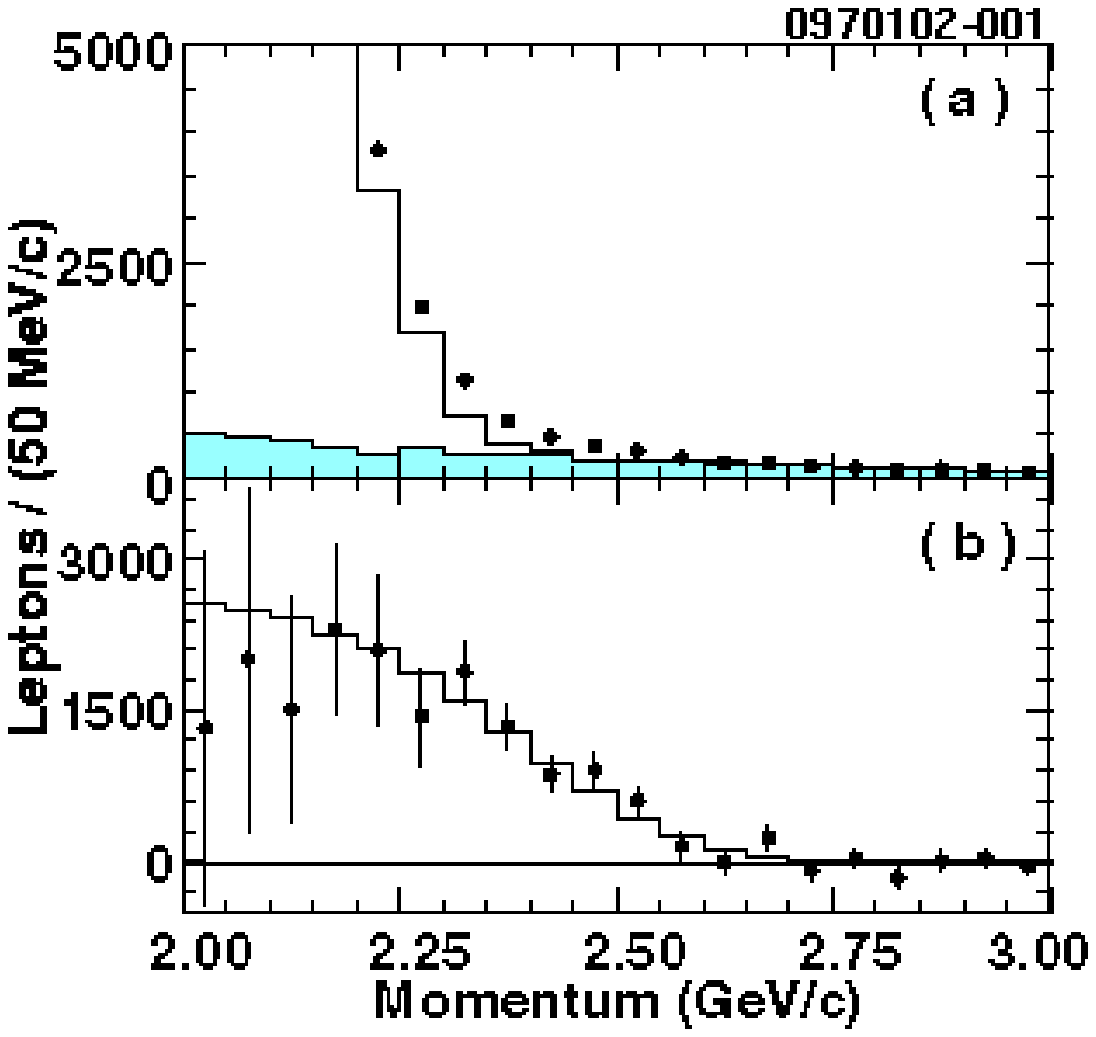,width=9cm}}
\begin{center}
\parbox{145mm}{\caption[ ]{\label{vub_incl.eps}\it {\bf (a)} Lepton spectra
for on-resonance data (points) and scaled off-resonance contributions
(shaded histo). The open histogram is the total background
(off-resonance + background B-decays).\\
{\bf (b)} Background-subtracted and efficiency corrected lepton spectrum for
$B\to X_u \ell \nu$ (points). The histogram is the $B\to X_u\ell\nu$
prediction based on the $B\to X_s\gamma$ spectrum.}}
\end{center}
\vspace*{-0.5cm}
\end{figure}

\paragraph{${\boldmath\CKM{ub}}$ from exclusive decays}
CLEO has updated the first $B \to \pi \ell \nu$ measurement
(\cite{vub_pilnu_old}) with improved statistics and event
reconstruction.  The larger data sample (CLEO Ii+II.5) allows us to extract 
signal rates in three independent
regions of the momentum transfer $q^2$. 
%The $q^2$ dependence of the form factor predictions allows us to
%reduce the systematic uncertainty on the form factor.
The separation into $q^2$ bins also permits the test of 
different form factor models and their $q^2$ dependence.  
The preliminary CLEO measurement \cite{vub_pilnu} of the branching
fraction
$\BR(B^0\to \pi^- \ell^+
\nu)=(1.376\pm0.180^{+0.116}_{-0.135}\pm0.008\pm0.102\pm0.021)\times
10^{-4}$ is based
on a form factor parameterization \cite{ball01} consistent with our
sub results in the $q^2$ bins. From \BR\ we derive a preliminary value of 
$|\CKM{ub}|=(3.25
\pm 0.21^{+0.16}_{-0.18}\:^{+0.64}_{-0.56}\pm0.12\pm0.07)\times10^{-3}$, 
where the uncertainties are statistical, systematic, and theory
uncertainties from the $\pi\ell^+\nu$ form factor, $\rho\ell^+\nu$ form
factor and from uncertainties due to other background from B decays.
We also obtain branching fractions for $B^0\to \rho^-\ell\nu$ and
$B^+\to \eta\ell^+\nu$. \cite{vub_pilnu}
\begin{figure}[htb]
\centerline{\epsfig{file=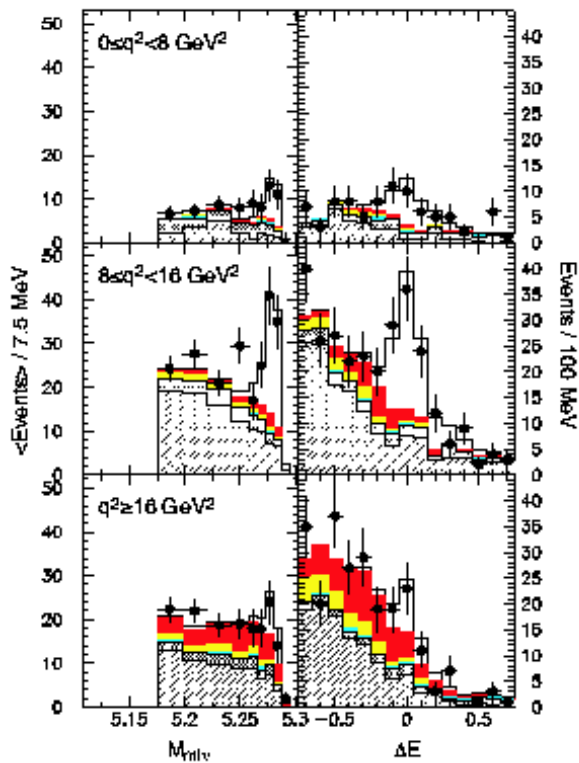,width=8cm}}
\begin{center}
\parbox{145mm}{\caption[ ]{\label{pi_q2_dq0.eps}\it  Reconstructed B
mass ($M_{m\ell\nu}$) and energy difference \deltae\ in the three $q^2$
regions for $B\to \pi \ell \nu$ 
%for the combined charged and neutral pion modes. 
Shown are on-resonance data (points), shaded histogram components are
background, open histo is signal. 
%are $b\to c$ (hatched), continuum (dotted), fake leptons (cyan or dark
%grey) feeddown from other $B\to X_u \ell\nu$ modes (yellow or light
%green), cross feed from other signal modes (red or black) and signal
%open.
}}
\end{center}
\vspace*{-0.5cm}
\end{figure}

The combination of quark mixing matrix results is
an on-going project. Common systematic effects and theoretical
uncertainties need to be taken into account. The resulting averages
have only slightly smaller total errors than individual
measurements\cite{pdg_vcb},\cite{pdg_vub}.
%CLEO has also another inclusive measurement employing neutrino
%reconstruction \cite{vub_neutrino}
The high statistics data samples accumulated by the B-factories in the
coming years will probably provide new measurements of $\CKM{cb}$ and
$\CKM{ub}$ using
fully reconstructed \yfours\ events\cite{ian_jik}.
%\clearpage
\section{Rare and Hadronic B decays}
\begin{figure}[htb]
\centerline{\epsfig{file=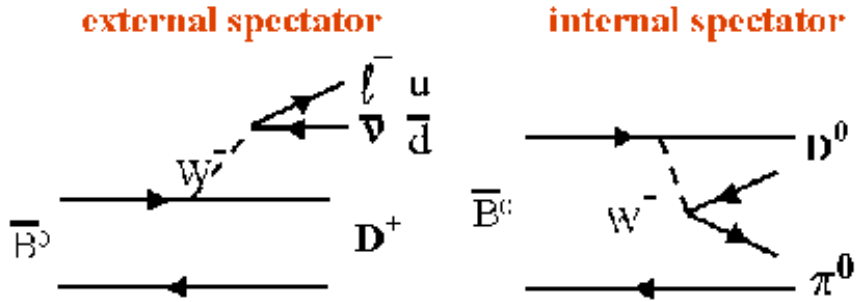,width=12cm}}
\begin{center}
\parbox{145mm}{\caption[ ]{\label{external_spec.eps}\it  
Example decay diagrams of B meson decays: external spectator, 
internal spectator diagram.}}
\end{center}
\vspace*{-0.5cm}
\end{figure}

The simplest B decay is the external spectator diagram given in
Fig. \ref{external_spec.eps}. In hadronic decays 
the internal spectator diagram is also possible. In the case of
$B^\pm$ decays this diagram can interfere with the external spectator
while it leads to unique final states in neutral B meson decays.

Phenomenological parameters $a_1$ and $a_2$ are
introduced to absorb non-perturbative contributions to the external
and internal spectator amplitudes, respectively. 
While theoretical results show that $a_1$ is 
process-independent\cite{neubert_beneke}, and one value is
sufficient to describe all decays, the process-independence of $a_2$
has no theoretical basis and experimental measurements are needed
here.

CLEO has dominated for a long time the measurement of exclusive B
decays as well as other areas of B physics. 
Comparing the CLEO measurements of hadronic B decays with the
new results from Belle and Babar, we find excellent\footnote{The branching
fraction for $B^- \rightarrow \phi K-$ might need further study. The
Belle and Babar measurements in the PDG 2002 edition are not quite in
agreement while the CLEO measurement is consistent with
both Belle and Babar.} agreement\cite{PDG2002}.

It is thus no surprise that 
measurements of exclusive hadronic B decays have reached
sufficient precision to challenge our understanding of the dynamics in
B decays. In analogy to semileptonic
decays, two-body hadronic decay amplitudes might be expressed as the product of
two independent hadronic currents, one describing the formation of a
charm meson and the other the transition of the virtual
W$^-$ into hadron(s). Considering the relatively large energy release
in B meson decays, the
$u\bar{d}$ pair, which is produced in a color singlet, travels fast
enough to leave the interaction region without influencing the second
hadron formed from the c quark and the spectator anti-quark. The
assumption that the amplitude can be expressed as the product of two
hadronic currents is called ``factorization''\cite{factorization}. 
This argument favors the external spectator diagrams. 

The internal spectator decay mode is suppressed compared to external
spectator processes, since the color of the quark-pair originating from the W 
decay must match the color of the other quark pair. 
In the decays of charm mesons, the effect of color-suppression is
present but final state interactions, or non-factorizable
contributions obscure its observation. The factorization is not as clear as
in the B meson system, due to the smaller momentum transfer in charm decays.
The concept of color suppression is, however,
much clearer in the B meson system. 

Until recently the
$B \to $ charmonium + X transitions were the only identified
color-suppressed B decays. 
CLEO\cite{bdpi0_cleo} and Belle\cite{bdpi0_belle} have recently observed the
color suppressed decays $\aBz \to \DDstz \gpz$.\footnote{Babar has also preliminary results for $\Btodpi$ \cite{bdpi0_babar}.}
The CLEO results are
$\BR(\Btodpi)$ = $({2.74}_{-0.32}^{+0.36}\pm0.55)\kreuz10^{-4}$, 
and $\BR(\Btodstpi)$ = $({2.20}_{-0.52}^{+0.59}\pm0.79)\kreuz10^{-4}$.

\begin{figure}[htb]
\centerline{\epsfig{file=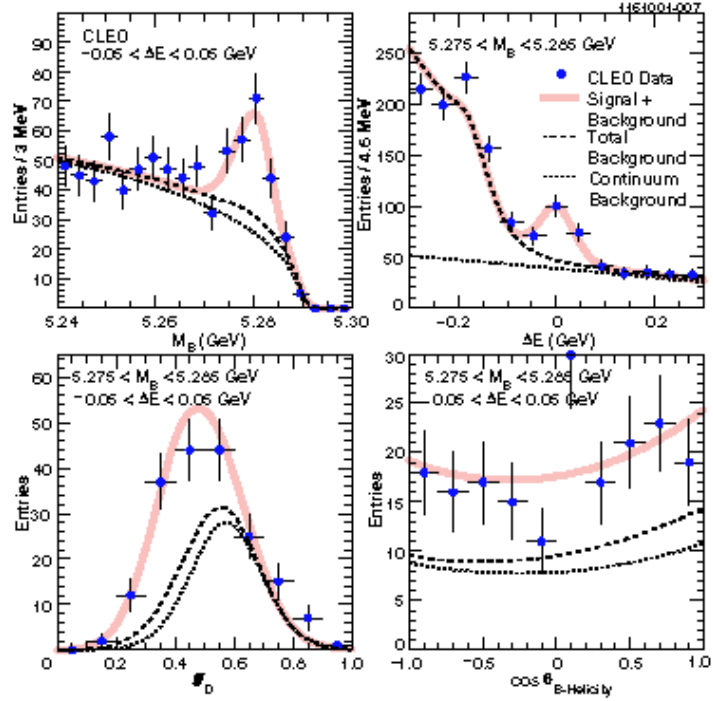,width=10.5cm}}
\begin{center}
\parbox{145mm}{\caption[ ]{\label{signalregion_overlay_bdpi0.eps}\it
$\aBz\to D^0\pi^0$.
The results of the unbinned, extended
maximum likelihood fit are shown as the 
full line. The dotted line represents the fitted 
continuum and the dashed line is the fit result for the sum of \BBar\ 
and continuum background. 
To enhance the signal for display purposes, the fit results are projected 
into the \mbc-\deltae\ signal region.}}
\end{center}
\vspace*{-0.3cm}\end{figure}

The signal yield is obtained from an
unbinned, extended maximum likelihood fit. The free parameters 
of the fit are
the number of signal events, background from B decays, and 
from continuum e$^+$e$^-$ annihilation.
Four variables are used as input to the maximum likelihood fit:
the beam-constrained mass \mbc, the energy difference \deltae,
the Fisher Discriminant \FD, which is a combination of event shape
variables, and the cosine of the decay angle of
the B \ctst, defined as the   
angle between the D momentum and the B flight direction 
calculated in the B rest frame.
The likelihood of the B candidate is the sum of
probabilities for the signal and two background hypotheses with
relative weights maximizing the likelihood.
Fig.~\ref{signalregion_overlay_bdpi0.eps} demonstrates 
the significance of our result.
Comparing our result to two-body B decays to charmonium the
process dependence of the phenomenological parameter $a_2$ is 
favored \cite{neubert_petrov_color_suppr}.

\begin{figure}[htb]
\centerline{\epsfig{file=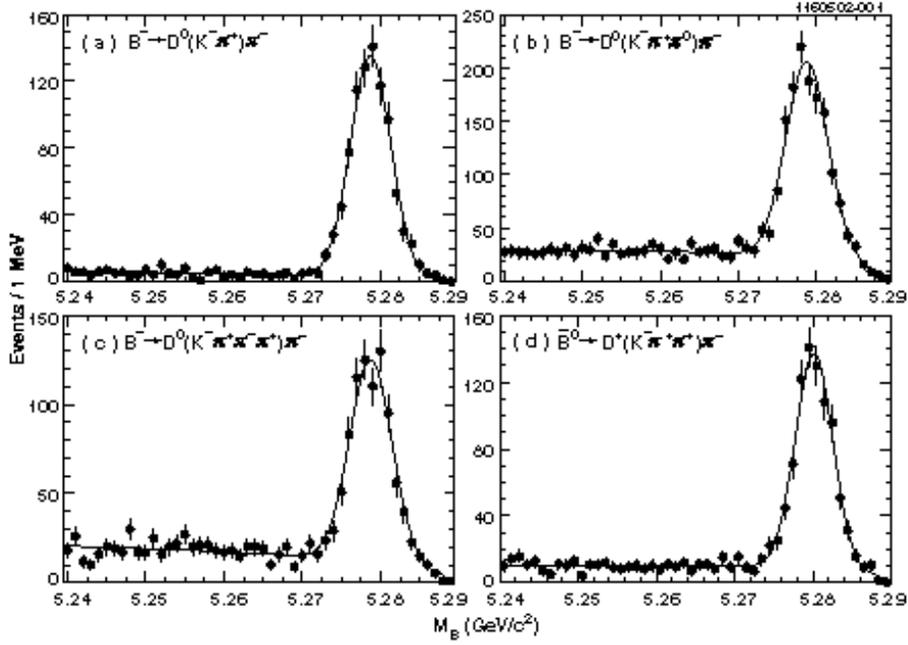,width=12.5cm}}
\begin{center}
\parbox{145mm}{\caption[ ]{\label{bdpi_figure.eps}\it The \mbc\
distributions for the $B\to D \pi$ candidates.}}
\end{center}
\vspace*{-0.5cm}
\end{figure}

The observation of \Btodpi\ completes the measurement of ${\rm
D\pi}$ final states and allows us perform an isospin analysis
and to extract the strong phase difference, $\delta_I$,
between isospin 1/2 and 3/2 amplitudes \cite{heavy_flavor,rosner99}. 
CLEO has improved its previous measurements of the color-favored 
$B \to D \pi$ decays\cite{bdpi}
\[ \BR(B^-\to D^0 \pi^-)=(49.7\pm 1.2\pm2.9\pm2.2)*10^{-4},\]
\[ \BR(\aBz\to D^+ \pi^-)=(26.8\pm1.2\pm2.4\pm1.2)*10^{-4},
\] where the errors are statistical, systematic and the error from the
uncertainty on the \yfours\ branching fraction.\cite{bdpi}
Because the error distribution of the phase $\delta_I$ is highly asymmetric
and non-Gaussian, we quote the cosine of the angle. We obtain
$\cos \delta_{I}= 0.863^{+0.024+0.036+0.038}_{-0.023-0.035-0.030}$
based on the CLEO color-favored (Fig.~\ref{bdpi_figure.eps}) and Belle's+CLEO's
color-suppressed results. 
The significance for the non-zero phase $\delta_I$ is
$2.3\sigma$ which suggests final state interactions. 
The fourth error on $\cos\delta_I$ is the uncertainty of
the \yfours\ branching fraction. The uncertainty on this basic
quantity affects significantly the
extraction of final state phases. The same is true for the extraction
of (weak) phases in $B \to \pi \pi$ once the signal yields are
measured with high enough precision. This has consequences for the
unitarity triangle since the extraction of the angle $\gamma$ relies
on the extraction of phases from the $B\to\pi\pi$ branching fractions.
The occurrence of final state interactions might also
obscure the extraction of $\gamma$ \cite{bdpi_gamma}.

%Models of hadronic B decay \cite{heavy_flavor} have successfully
%described experimental results using two phenomenological parameters,
%$a_1$ and $a_2$, that characterize non-factorizable contributions. Both are
%believed to be process-dependent but so far experimental data have
%been consistent with universal values for $a_1$ and $a_2$.
%Recent work by Beneke, Buchalla, Neubert and Sachrajda \cite{neubert_beneke} 
%has shown that $a_1$ is only slightly process-dependent.
%Comparing our result to two-body B decays to charmonium the
%process dependence of $a_2$ is favored \cite{neubert_petrov_color_suppr}.
%difference in lifetime between D0 and D+, due to sestructive
%interference between internal, external spectator diagram. while
%partial leptonic widths are the same.  smaller diff observed in B0
%B+. 
\paragraph{${\boldmath B \to K \pi \pi}$}
CLEO measurement of charm-less hadronic two-body decays
have received
considerable attention because of their importance for unitarity
triangle measurements. A natural extension of these measurements are 
three-body modes. These modes might reveal two-body channels
with intermediate vector resonances which provide complementary
information for the unitarity triangle measurements. 

CLEO analyzed $K^0_s h^+ pi^-$, $K^+h^-\pi^0$ 
and  $K^0_s h^+\pi^0$, where $h_\pm$ denotes a charged pion or kaon\cite{bkpipi}.
Obvious contributions from B decays into charm
are removed from our sample by cuts on the invariant masses, namely, 
$B \to D \pi$, $D\to K \pi$ in addition to $B \to \JPsi K^0$, $\JPsi \to \mu^+
\mu^-$, where the muons are misidentified as pions. 
Signal yields are extracted from unbinned maximum likelihood fits with
several Dalitz contributions. Interference between these amplitudes is
neglected and taken into account as a systematic uncertainty. 

\begin{figure}[htb]
\ifnohepex\vspace*{2cm}\fi
\centerline{\hspace*{-0.4cm}\epsfig{file=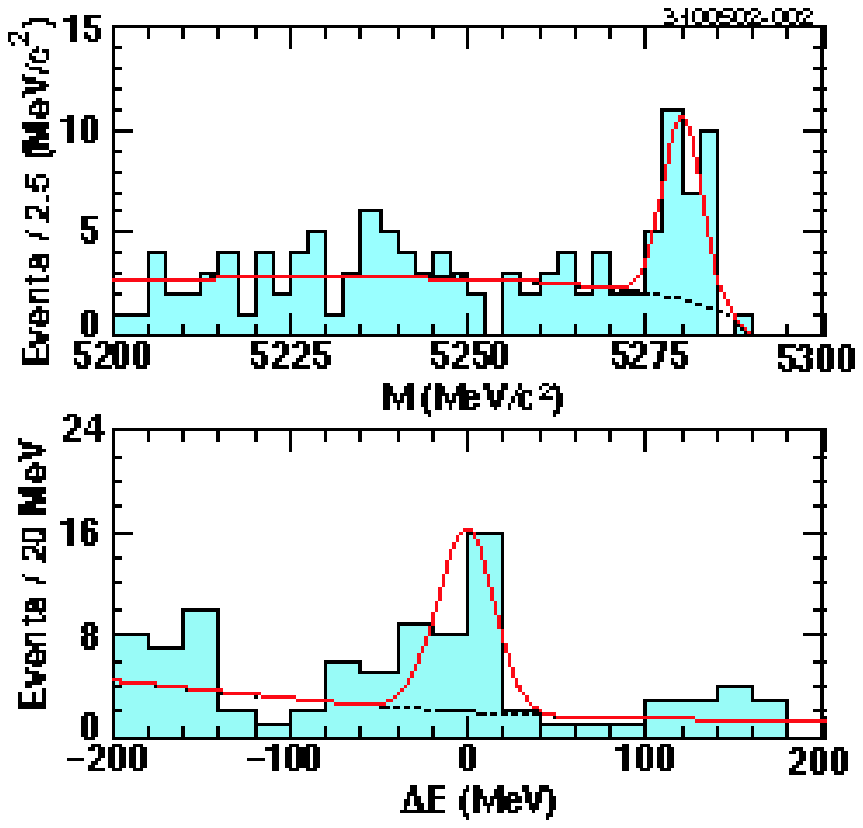,width=7.0cm}\hspace*{-0.5cm}\epsfig{file=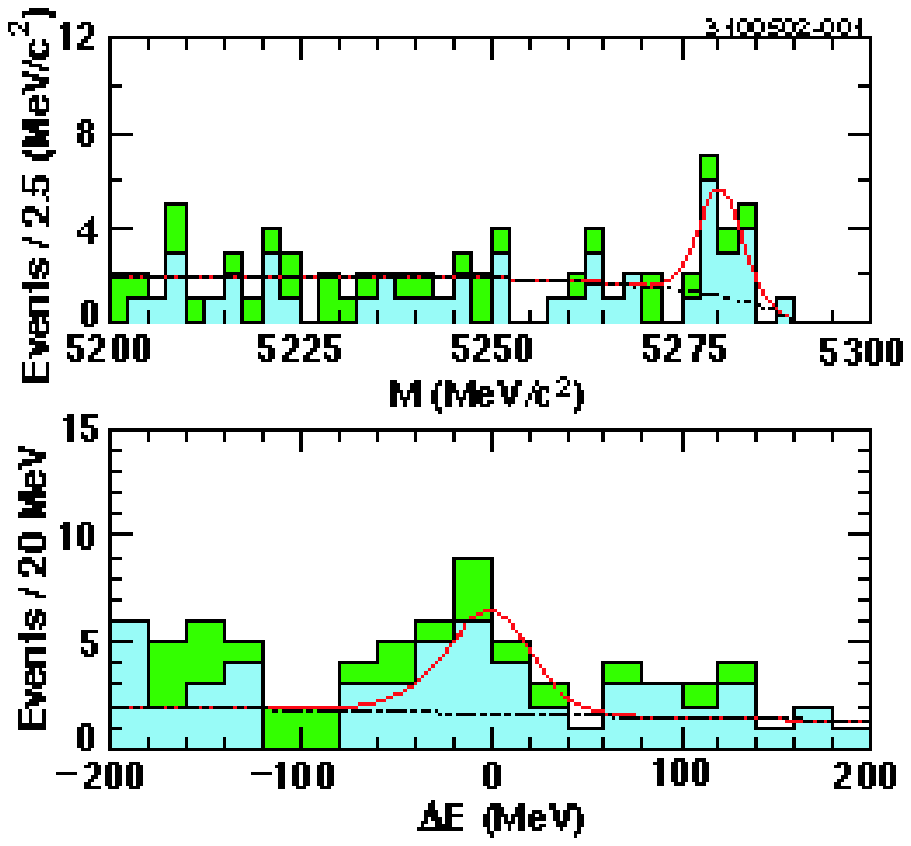,width=7.5cm}}
\begin{center}
\parbox{145mm}{\caption[ ]{\label{3100502-001.eps}\it  M and \deltae\
projections for $B \to K^0 \pi^+ \pi^-$ (left), and $B \to \Kstpl
\pi^-$ (right). The latter includes the two \Kstpl\ submodes,
$\Kstpl\to K^0\pi^+$ (light shade) and $\Kstpl\to K^+\pi^0$ (dark
shade). The background has been suppressed in the plot by a cut
on event probabilities. Fit
results for background (dashed line) and signal+background (full line)
are also shown.}}
\end{center}
\vspace*{-0.5cm}
\end{figure}

\begin{figure}[htb]
\centerline{\epsfig{file=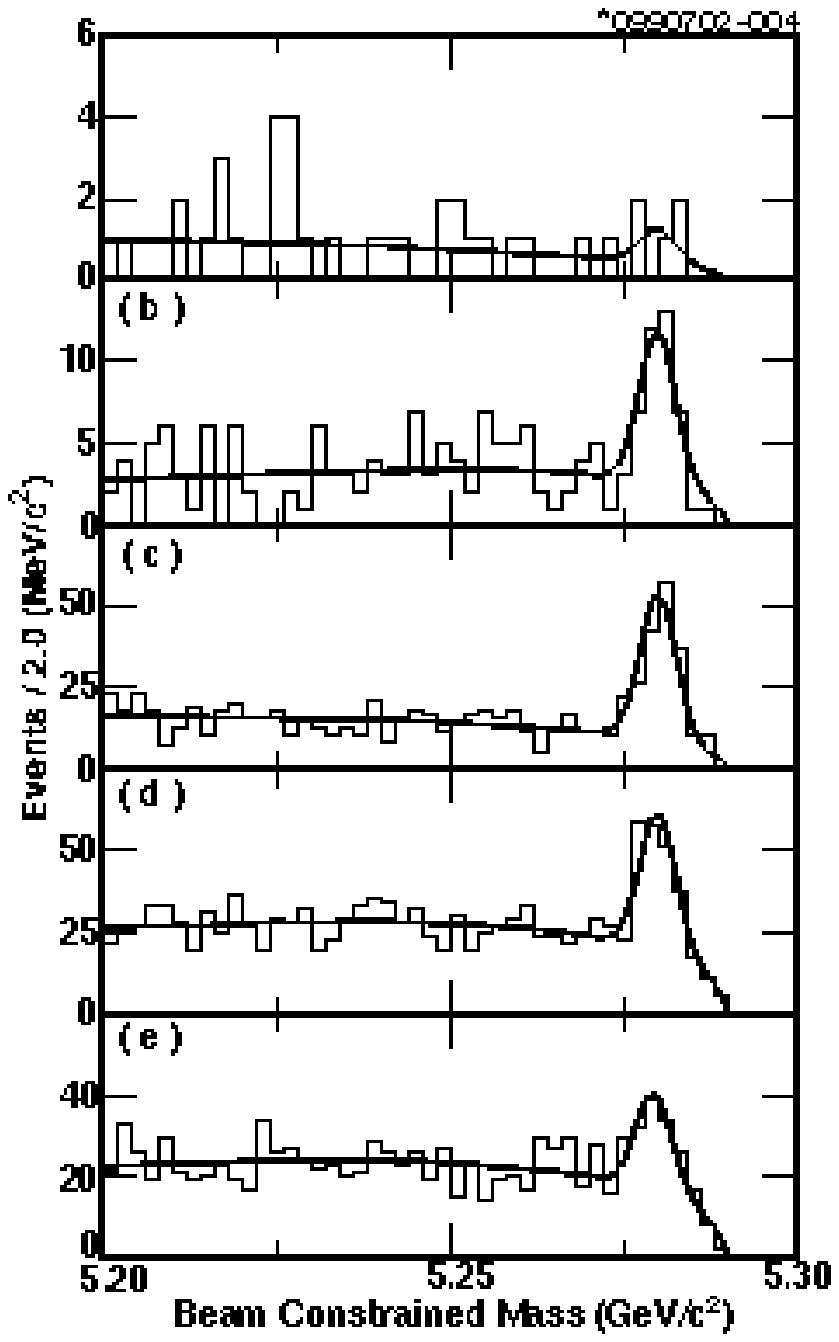,width=7cm}\epsfig{file=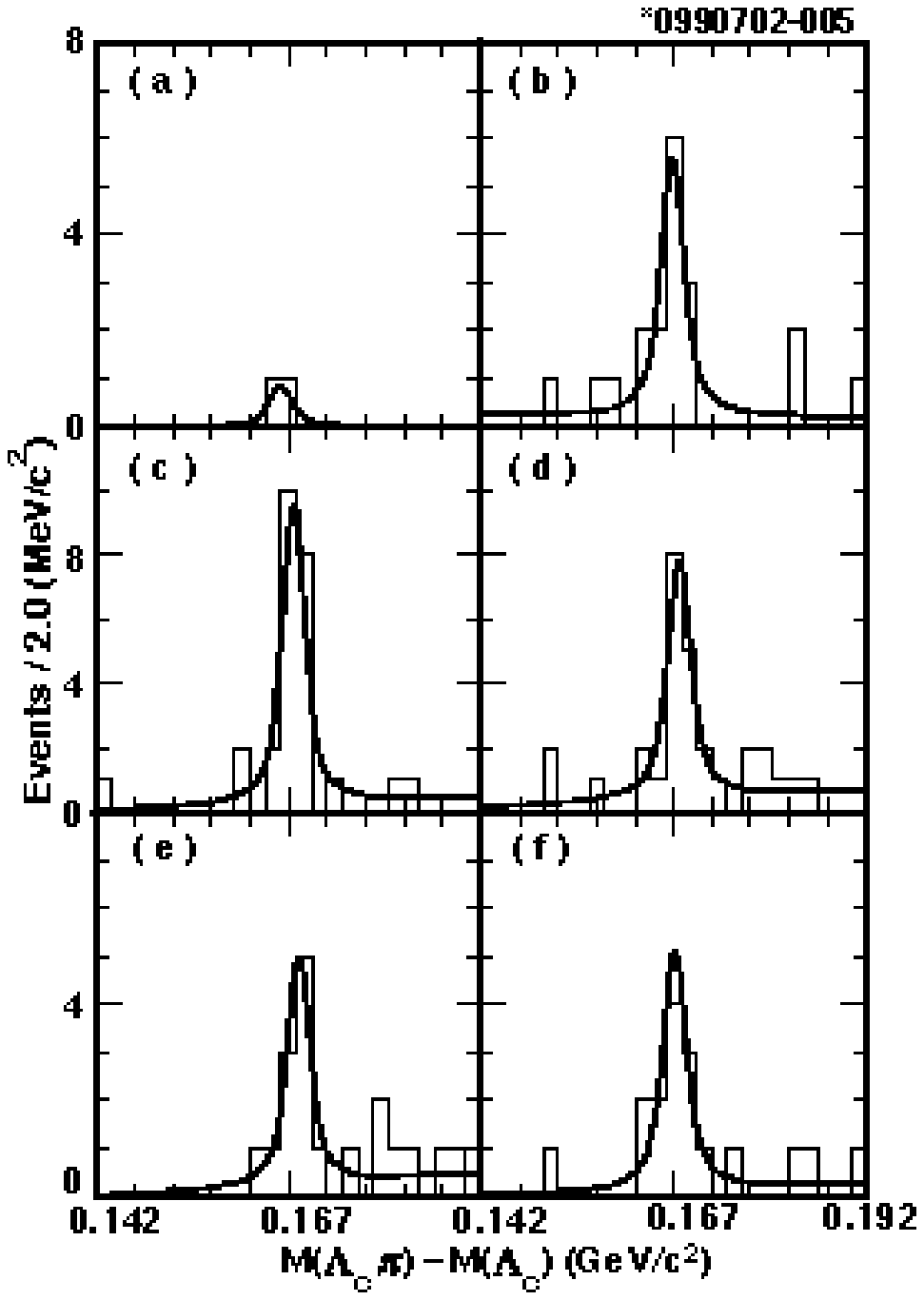,width=8cm}}
\begin{center}
\parbox{145mm}{\caption[ ]{\label{blcpfigure1.eps}\it left side: Beam
constrained mass distributions for a) $\lamc\bar{p}$, b)
$\lamc\bar{p}\pi^-$, c) $\lamc\bar{p}\pi^-\pi^+$, d)
$\lamc\bar{p}\pi^-\pi^+\pi^-$, e) $\lamc\bar{p}\pi^-\pi^0$\\
right side: $M(\lamc\pi)-M(\lamc)$ mass differences. a)
$\lamc\bar{p}\pi^-$, b) $\lamc\pi^-$ within $\lamc\bar{p}\pi^-\pi^+$,
c) $\lamc\pi^+$ within $\lamc\bar{p}\pi^-\pi^+$, d) $\lamc\pi^-$
within $\lamc\bar{p}\pi^-\pi^+\pi^-$ both combinations, e) $\lamc\pi^+$
within $\lamc\bar{p}\pi^-\pi^+\pi^-$, f) $\lamc\pi^-$
within $\lamc\bar{p}\pi^-\pi^0$.
}}
\end{center}
\vspace*{-0.5cm}
\end{figure}

We derive limits between 19 and 66 $\times 10^{-6}$ for five 
decay modes and observe $B \to K^0 \pi^+ \pi^-$ with a branching
fraction of $\BR = (50^{+10}_{-9}(stat)\pm7(syst)) \times 10^{-6}$. 
We perform Dalitz plot fits to search for a substructure
and find a contribution from $B \to \Kstpl \pi^-$. Since this mode
contributes also to $B \to K^+ \pi^0 \pi^-$ via $\Kstpl \to$ 
$K^+ \pi^0$, we fit these two modes simultaneously. 
The branching fraction is $\BR(B \to \Kstpl \pi^-)$ = $(16^{+6}_{-5}\pm2) \times 10^{-6}$ and
the signal is 4.6 $\sigma$ significant.
Our results are shown in Fig.~\ref{3100502-001.eps}.
%\clearpage
\paragraph{Baryonic B decays}
Decays of B hadrons into final states containing a
baryon-antibaryon pair have been known for some time. The inclusive
rate of $B \to \lamc + X$ is about 5\%, much larger than the sum of
exclusive decay modes. This suggest significant contributions from
final states containing a baryon-antibaryon pair and multiple pions.
We report new measurements \cite{blcp} 
of exclusive decays of B mesons into final
states of the type $\lamc \bar{p} n(\pi)$, where $n=0,1,2,3$. We find
signals in modes with 1, 2 and 3 charged pions and we derive an
upper limit for 
the two-body decay into $\lamc \bar{p}$. Our measurements are in good
agreement with our old results \cite{blcp_old}.
We obtain the branching fractions given in Table \ref{blcp.tab}. 
The beam-constrained mass of these decay modes is given in
Fig.~\ref{blcpfigure1.eps}, left side. 
We derive only a limit on the
simplest two-body decay mode $B \to \lamc \bar{p}$. Our limit is in
agreement with the recent observation of this mode by Belle \cite{lc_p_obs}.

\begin{small}
\begin{table}[htb]
\begin{center}
\begin{tabular}{lcc}
\hline\hline
Mode     & ${\cal B}$ ($10^{-4}$) & 
Previous Result ($10^{-4}$)\cite{blcp_old} \\
\hline
\hline
$\Lambda_c^+\overline{p}$      & $<0.9$                    & $<2.1$  \\
\hline
$\Lambda_c^+\overline{p}\pi^-$ & $2.4\pm0.6^{+0.19}_{-0.17}\pm0.6$ &
$6\pm3$ \\
\ \ \ \ \ $\Sigma_c^0\overline{p}$   & $<0.8$&\\
\hline
$\Lambda_c^+\overline{p}\pi^-\pi^+$        &$16.7\pm1.9^{+1.9}_{-1.6}\pm
4.3$ &$13\pm6$ \\
 \ \ \ \ \ $\Sigma_c^0\overline{p}\pi^+$ & $2.2\pm0.6\pm0.4\pm0.6$ &  \\
 \ \ \ \ \ $\Sigma_c^{++}\overline{p}\pi^-$ & $3.7\pm0.8\pm0.7\pm1.0$& \\
 \ \ \ \ \ $\Lambda_{c1}^+\overline{p}$   &    $<1.1$                &\\
\hline
$\Lambda_c^+\overline{p}\pi^-\pi^+\pi^-$ & $22.5\pm2.5^{+2.4}_{-1.9}\pm5.8
$   & $<15$ \\
 \ \ \ \ \ $\Sigma_c^0\overline{p}\pi^+\pi^-$ & $4.4\pm1.2\pm0.5\pm1.1$ & \\
 \ \ \ \ \ $\Sigma_c^{++}\overline{p}\pi^-\pi^-$ & $2.8\pm0.9\pm0.5\pm0.7$ &  \\
 \ \ \ \ \ $\Lambda_{c1}^+\overline{p}\pi^-$   &    $<1.9$                &\\
\hline
$\Lambda_c^+\overline{p}\pi^-\pi^0$ & $18.1\pm2.9^{+2.2}_{-1.6}\pm4.7$ & $
<31$ \\
 \ \ \ \ \ $\Sigma_c^0\overline{p}\pi^0$ & $4.2\pm1.3\pm0.4\pm1.1$ &  \\
\hline\hline
\end{tabular}
\parbox{145mm}{\caption[ ]{\label{blcp.tab}\it Branching fractions or 90\% C.L. upper
limits from CLEO\cite{blcp} compared to our old results
\cite{blcp_old}. Substructure results are given in the indented rows. 
The second error in the branching
fraction is due to all systematic uncertainties except for the
uncertainty due to the measurement of the $\Lambda_c^+\to pK^-\pi^+$ 
branching fraction, which is kept separate and appears as a third
uncertainty.}}
  \end{center}
\end{table} 
  \end{small}

The $\lamc$ and one of the pions might come from higher resonances. We
have searched for a substructure in the various Dalitz decay
plots. Fig.~\ref{blcpfigure1.eps}, right side, shows the distribution
of the $\lamc \pi
-\lamc$ mass difference in the vicinity of the $\Sigma_c$
resonances. Utilizing CLEO's precise mass and width measurements of these
resonances \cite{sigmac}, we are able to estimate the significance of
our signals and derive branching fractions for several modes (see Table
\ref{blcp.tab}). 

Again, we have {\bf not} observed true two-body decay
modes (of the form $B \to \Sigma_c \bar{p}$). 
Our newly observed three-body decay modes $\aBz\to\Sigma_c^{++}
\bar{p}\pi^-$, $\aBz\to\Sigma_c^{0}\bar{p}\pi^+$, $\aBz\to\Sigma_c^{0}
\bar{p}\pi^0$ have essentially identical phase space, but only the
$\Sigma_c^{++}$ decay can proceed via both external and internal
spectator diagrams, whereas the $\Sigma_c^{0}$ decay can only proceed
via an internal spectator diagram. We find the the rate of all three
decay modes to be of the same order. This implies that the external W
decay diagram does not dominate over the internal spectator, although
naively we would expect the latter to be color-suppressed.
 
The large discrepancy compared to color-suppressed B decays into mesons 
might be explained by the smaller momentum transfer in baryonic B
decays due to the larger mass of the baryon-antibaryon system. 
%although a simple calculation shows that  
%the maximum momentum transfer to the \pi^- in $B \to \lamc \bar{p}
%\pi^-$ (1.6 GeV) is larger than a typical momentum transfer in D
%decays. (about 0.8 GeV) kumac twobody_p m= m1= m2=
A different explanation is given in \cite{blcp_expl}.

\clearpage
\section{\ythrees\ Spectroscopy}
The spectroscopy of bound $b\bar{b}$ states is an excellent testing
ground for lattice QCD \cite{lepage_ssi}. The spin-triplet S-wave
states $\Upsilon(nS)$ with $J^{PC}=1^{--}$ are produced in $e^+e^-$
annihilation. These states can decay radiatively with an electric
dipole transition (E1) to the spin-triplet P-wave levels, $\chi_b(nP_J)$. 
Subsequent decays can either return to a lower $\Upsilon(nS)$ state, to the
spin-singlet S-wave states $\eta_b(nS)$ or the $\Upsilon(nD)$ states. 
Neither of the $\eta_b(nS)$ or the $\Upsilon(nD)$ state had been observed by CLEO-II, ARGUS or CUSP.
CLEO-III has accumulated new data sets on the $\Upsilon(1S)$,
$\Upsilon(2S)$ and $\Upsilon(3S)$, that are comparable in size or
larger than the existing
world data sets. The first set to become available for data analysis
were 4.73 Million $\Upsilon(3S)$ decays collected with the CLEO-III
detector. The CLEO-III sample constitutes roughly a ten-fold increase
in $\Upsilon(3S)$ statistics compared to the CLEO-II data set
\cite{cleotwo_ups3s}. 
\begin{figure}[htb]
\centerline{\epsfig{file=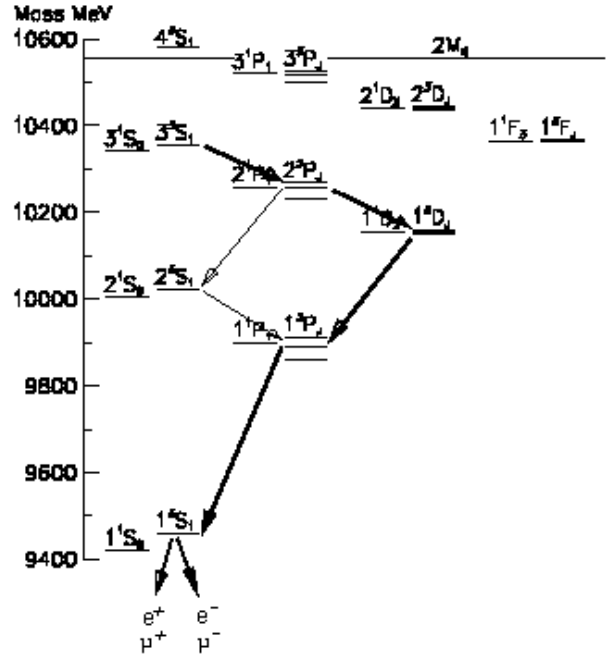,width=9cm}}
\begin{center}
\parbox{145mm}{\caption[ ]{\label{mass_levels_1d_and_2s}\it Mass
spectrum of bound $b\bar{b}$ states.}}
\end{center}
\vspace*{-0.5cm}
\end{figure}

\paragraph{Search for the ${\boldmath\eta_b(1S)}$.}
The $\eta_b(1S)$ is the ground state of the $b\bar{b}$ system.
\begin{figure}[htb]
\centerline{\epsfig{file=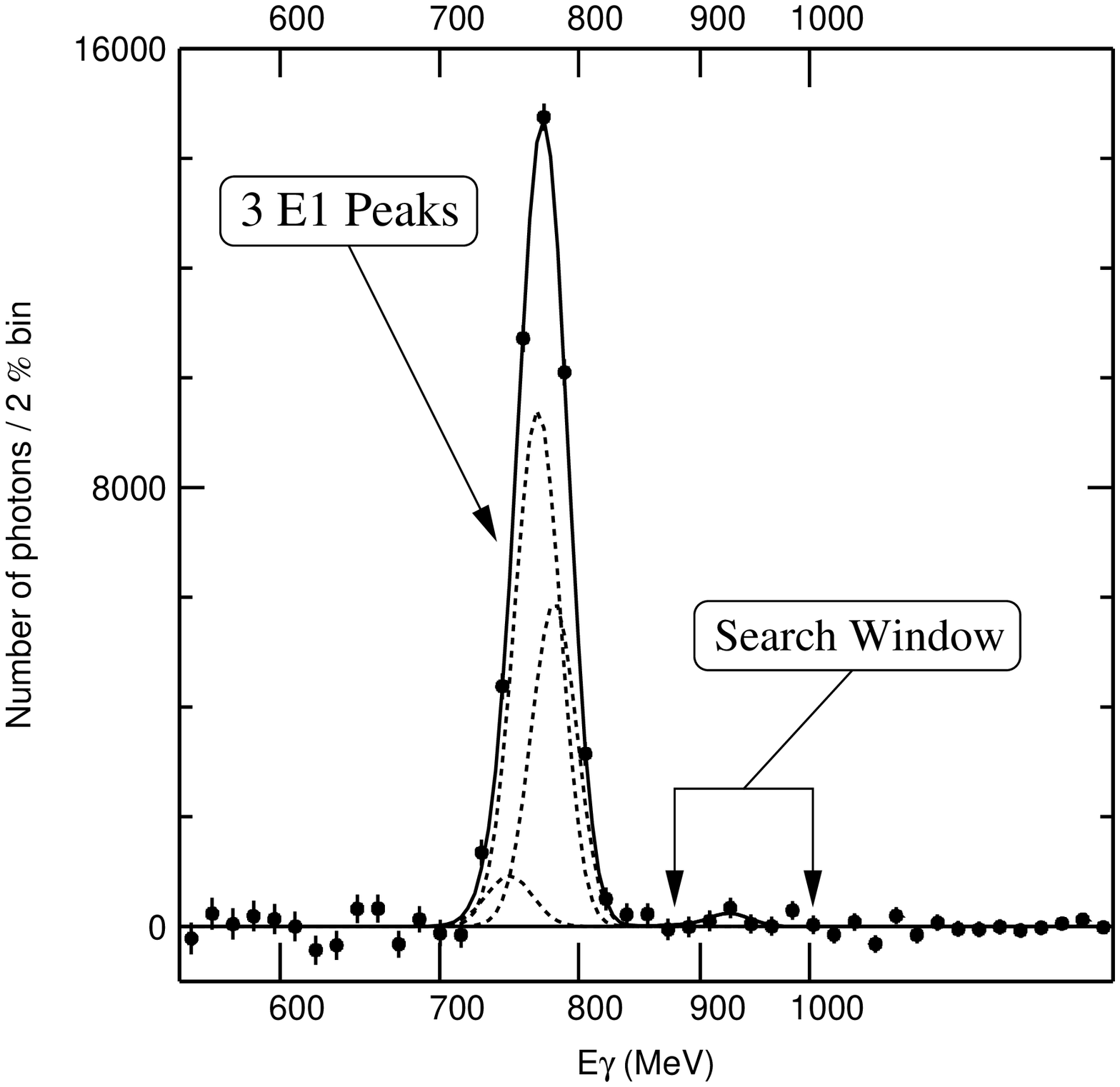,width=8cm}\epsfig{file=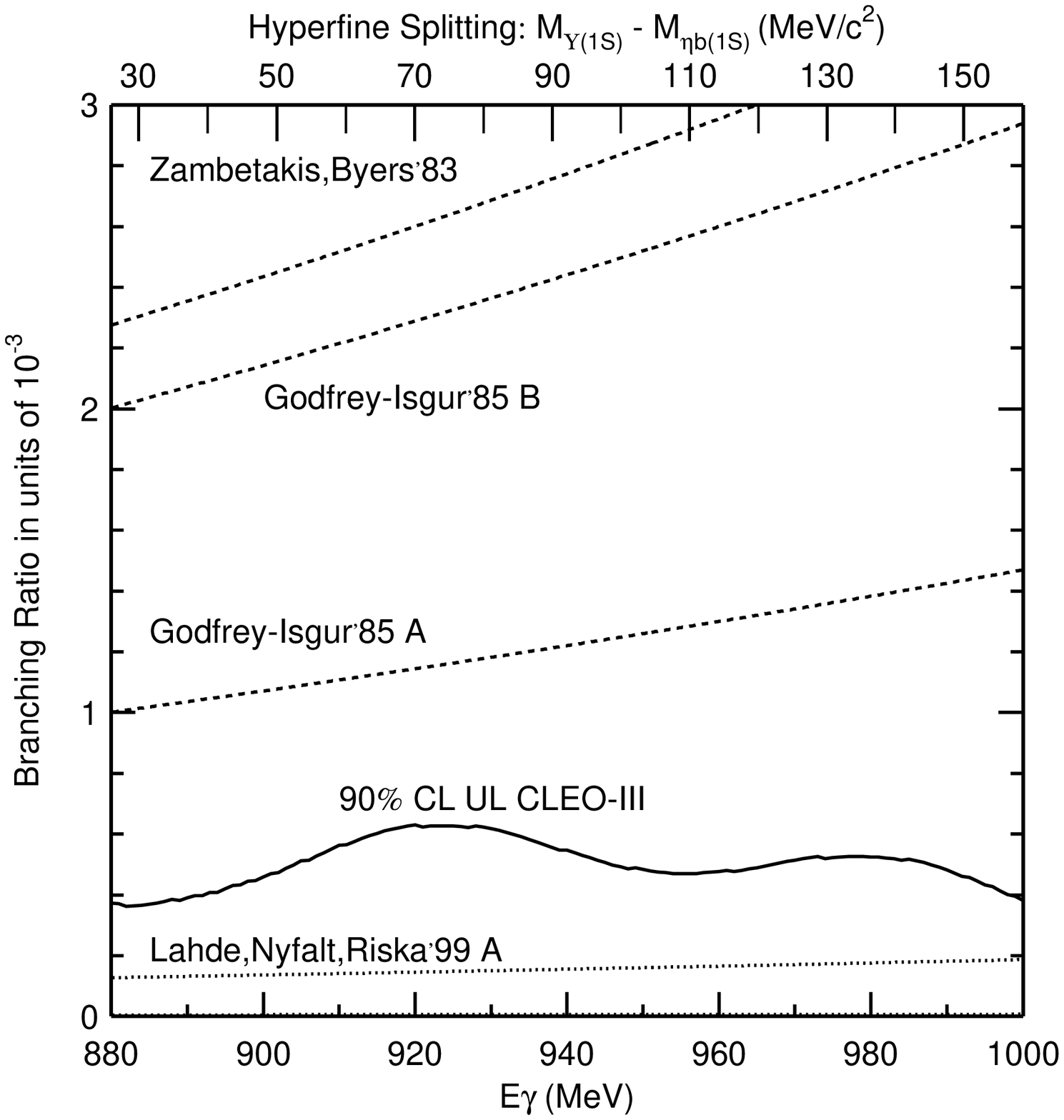,width=8cm}}
\begin{center}
\parbox{145mm}{\caption[ ]{\label{search_etab.eps}\it {\bf (left)} Background
subtracted photon spectrum in the $\chi_b(2P_J)\to \gamma
\Upsilon(1S)$ region (\tilde 780 MeV) and the search window. The
background was subtracted with a polynomial fit plus a Gaussian for
the E1 peak.
{\bf (right)} Preliminary upper limits on $\BR(\Upsilon(3S)\to\eta_b(1S)\gamma)$
with 90\% confidence level. Predictions are taken from \cite{etabprediction}.}}
\end{center}
\vspace*{-0.5cm}
\end{figure}
To reach the $\eta_b(1S)$, it is necessary to detect either 
favored magnetic dipole transitions (M1) with very small photon
energies or hindered M1 transitions with changes in the principal
quantum number. Since there are abundant exclusive decay modes
of the \etab \footnote{So far only one search for the $\eta_b(1S)$
via exclusive
decay modes has been proposed \cite{braaten_etab}, 
utilizing the expected branching fraction
$\eta_b(1S)\to\Jpsi\Jpsi$ of order $7\times 10^{-5} - 7\times 10^{-3}$. This strategy might be
applicable at the Tevatron.}
an inclusive search strategy is the most promising approach. Since the
M1 transition $\Upsilon(1S)\to\eta_b(1S)\gamma$ is suffering from a
small phase space and a huge low-energy photon background, the
hindered M1 transition is a promising decay mode.

We analyze the inclusive photon spectrum of well-contained hadronic
events on the $\Upsilon(3S)$ resonance. The range of theoretical
predictions of the $\eta_b(1S)$-mass define a search window that
corresponds to photon energies between 880 to 1000 MeV.
In this energy range the largest background arises from photons from
$\gpz$ decay. We reject photons that can be paired with another photon
to form a $\gpz$ candidate. The sensitivity of our search can be
investigated with a peak in the photon energy spectrum due to
$\chi_b(2P_J)\to \gamma \Upsilon(1S)$ transitions. This peak is at around 780 MeV
-- well below our search window. Figure \ref{search_etab.eps} shows the
background-subtracted photon spectrum. The peak for $\chi_b(2P_J)\to
\gamma \Upsilon(1S)$ demonstrates our good sensitivity to photons from
radiative transitions. We perform a series of fits
with a Gaussian signal shape assuming several peak energies,
obtaining a maximum signal yield of $698\pm 463$ which is only $1.5\:
\sigma$ significant. Since we do not find evidence for a signal, we
set upper limits on $\BR(\Upsilon(1S)\to\eta_b(1S)\gamma)$. The
preliminary $\BR$-limits as function of the photon energy $E_\gamma$
are shown in Fig.~\ref{search_etab.eps}, right side. We exclude most model
predictions \cite{etabprediction} with a C.L. of 90\% or better.\cite{etabsearch}
\paragraph{Two-photon cascades of the ${\boldmath\Upsilon(3S)}$.}
CLEO has updated its analysis of the $\chi_b(2P_J)$ states in an
analysis of the cascade decay $\Upsilon(3S)\to\chi_b(2P_J)\gamma$ ;
$\chi_b(2P_J)\to \gamma \Upsilon(nS)$ ; $\Upsilon(nS)\to
\ell^+\ell^-$, with n=1,2. We obtain new, preliminary 
mass measurements \cite{twogammatrans}
\[
m(\chi_b(2P_2)) = (10268.75 \pm 0.30(stat.) \pm 0.58(syst.)) MeV
\]
\[ 
m(\chi_b(2P_1)) = (10255.64 \pm 0.17(stat.) \pm 0.60(syst.)) MeV
\]
and also updated our previous branching fraction results.
In addition these measurements are an important cross-check of
multi-photon cascades, the four-photon cascades probably being 
the most interesting.
\paragraph{Discovery of the ${\boldmath\Upsilon(1D)}$.}
Recent interest in quarkonium spectroscopy arises from the possibility that our
measurements will aid theorists in understanding heavy quarkonium from
first principles QCD, given that there is a wide variety of the
spin-dependent splittings predicted by several calculations. 
The discovery of new $b\bar{b}$ states would
pose an important test. One proposed search
strategy\cite{godfrey_rosner_y1d} for the D-wave
state is via four-photon cascades from the $\Upsilon(3S)$ down to
the $\Upsilon(1S)$.
The signature of four-photon cascades can stem
from several different sources.
\begin{itemize} 
\item Photon cascade via the $\Upsilon(2S)$:\\ $\Upsilon(3S) \to \chi_b(2P_J) (+\gamma)\to \Upsilon(2S)
(+\gamma)\to \chi_b(1P_J)(+\gamma) \to \Upsilon(1S)(+\gamma)$
\item Hadronic transition\\ $\Upsilon(3S) \to \pi^0\pi^0\Upsilon(1S)$
\item Photon cascade via the $\Upsilon(1D)$:\\ $\Upsilon(3S) \to \chi_b(2P_J) (+\gamma)\to \Upsilon(1D)
(+\gamma)\to \chi_b(1P_J)(+\gamma) \to \Upsilon(1S)(+\gamma)$
\end{itemize}
The latter source is the signal we are looking for. 
CLEO has made the first observation of the $\Upsilon(1D)$
with these four-photon cascades \cite{upsoned}. Requiring that the endpoint of the
photon cascade, the $\Upsilon(1S)$, decays into a pair of leptons, we
have a clean signature of $4\gamma\ell^+\ell^-$. 
Cascades compatible with hadronic transitions $\Upsilon(3S) \to
\pi^0\pi^0\Upsilon(1S)$ or four-photon cascades via the $\Upsilon(2S)$
are vetoed. Due to an unfortunate combination of spin constraints, 
the latter veto also removes the
largest part of the expected $\Upsilon(1D_{J=3})$ signal 
for most of the mass range, leaving us sensitive to two out of
three $\Upsilon(1D)$ states.
 
The kinematics of the signal cascade can easily be reconstructed once the
photons have been assigned to the correct part of the decay chain. Two
of the four photon energies in the 
cascade are known, namely $\Upsilon(3S) \to \chi_b(2P_J)+\gamma$ and
$\chi_b(1P_J)\to\Upsilon(1S)+\gamma$. The other two energies depend on
the mass of the $\Upsilon(1D)$. Uncertainties in the energy
measurements increase the difficulty of finding the correct
assignment. For each possible assignment of the four photons to the
cascade we define a chi-square
 
\[ \chi^2_{1D,J_2P,J_1P} = \sum^{4}_{j=1}\left(\frac{E_{\gamma
j}-E_{\gamma j}^{expected}(M_{\Upsilon(1D)})}{\sigma_{E_{\gamma
j}}}\right)^2,\]
where $E_{\gamma j}$ are the measured photon energies and $E_{\gamma
j}^{expected}$ are the expected photon energies from the masses of the
$b\bar{b}$ states and the measured photon directions in the cascade.
The $\chi^2_{1D,J_2P,J_1P}$ 
depends on the assumed $\Upsilon(1D)$ mass and on the choice of
intermediate $J_{2P}$ and $J_{1P}$ states.  
We assign to each event an $\Upsilon(1D)$ 
candidate mass, $m(1D)$, which is the mass that minimizes
$\chi^2_{1D,J_2P,J_1P}$, trying all possible photon and spin combinations.

\begin{figure}[p]
\centerline{\epsfig{file=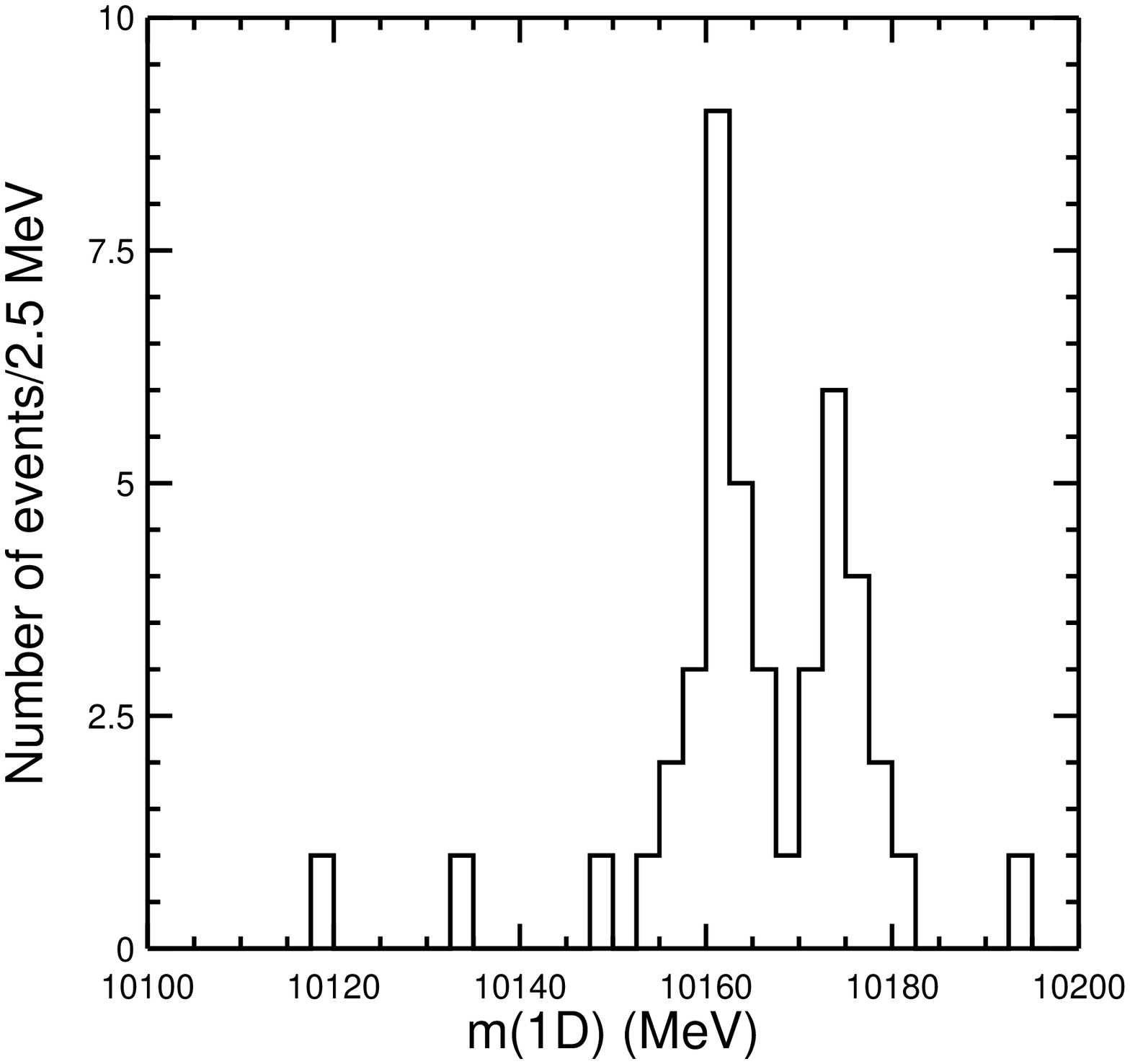,width=8cm}\epsfig{file=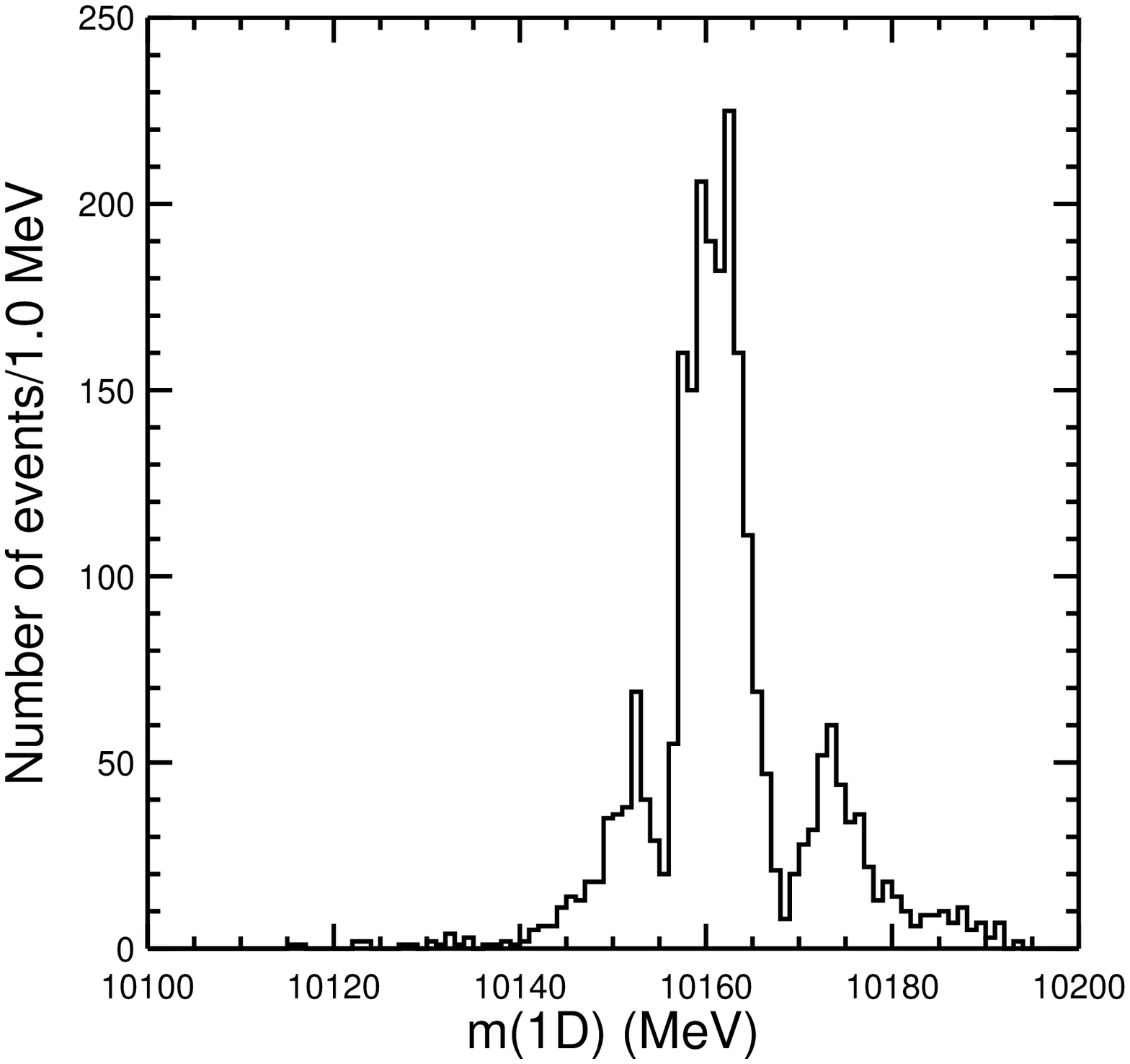,width=8cm}}
\begin{center}
\parbox{145mm}{\caption[ ]{\label{casc_12_m1df_1d2.eps}\it {\bf left
side:} Distributions of the most likely mass assignment $m(1D)$. {\bf right
side:} Monte Carlo
simulation of the reconstructed mass m(1D) for an $\Upsilon(1D)$ state
of M=10160 MeV.}}
\end{center}
\vspace*{-0.5cm}
\end{figure}

Distributions of the most likely mass assignment $m(1D)$ is shown in 
Fig.~\ref{casc_12_m1df_1d2.eps}, left side. From our Monte Carlo
simulations we expect to see signal mass peaks with smaller satellite peaks
as shown in Fig.~\ref{casc_12_m1df_1d2.eps}, right side. 
\begin{figure}[p]
\centerline{\epsfig{file=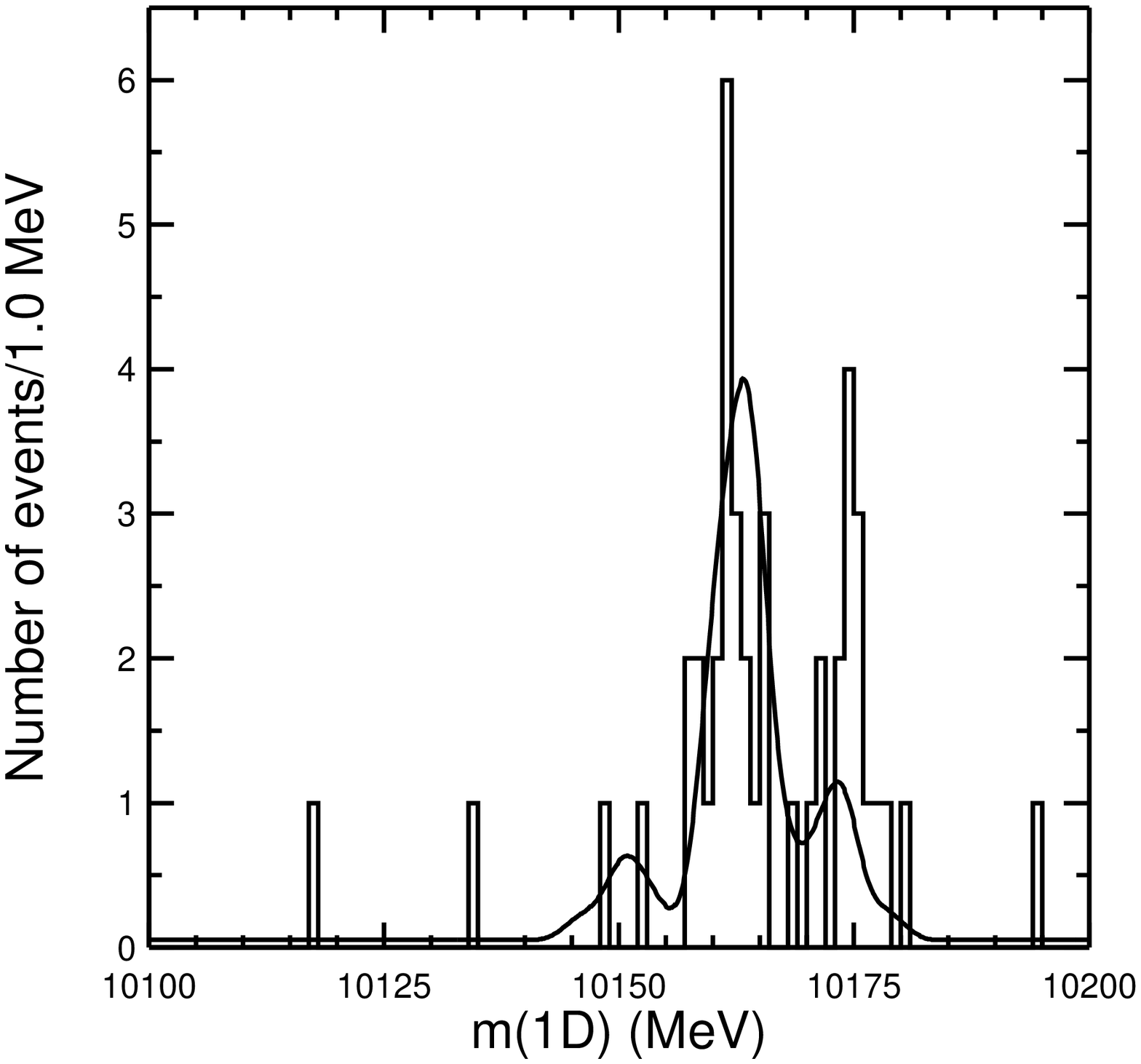,width=8cm}\epsfig{file=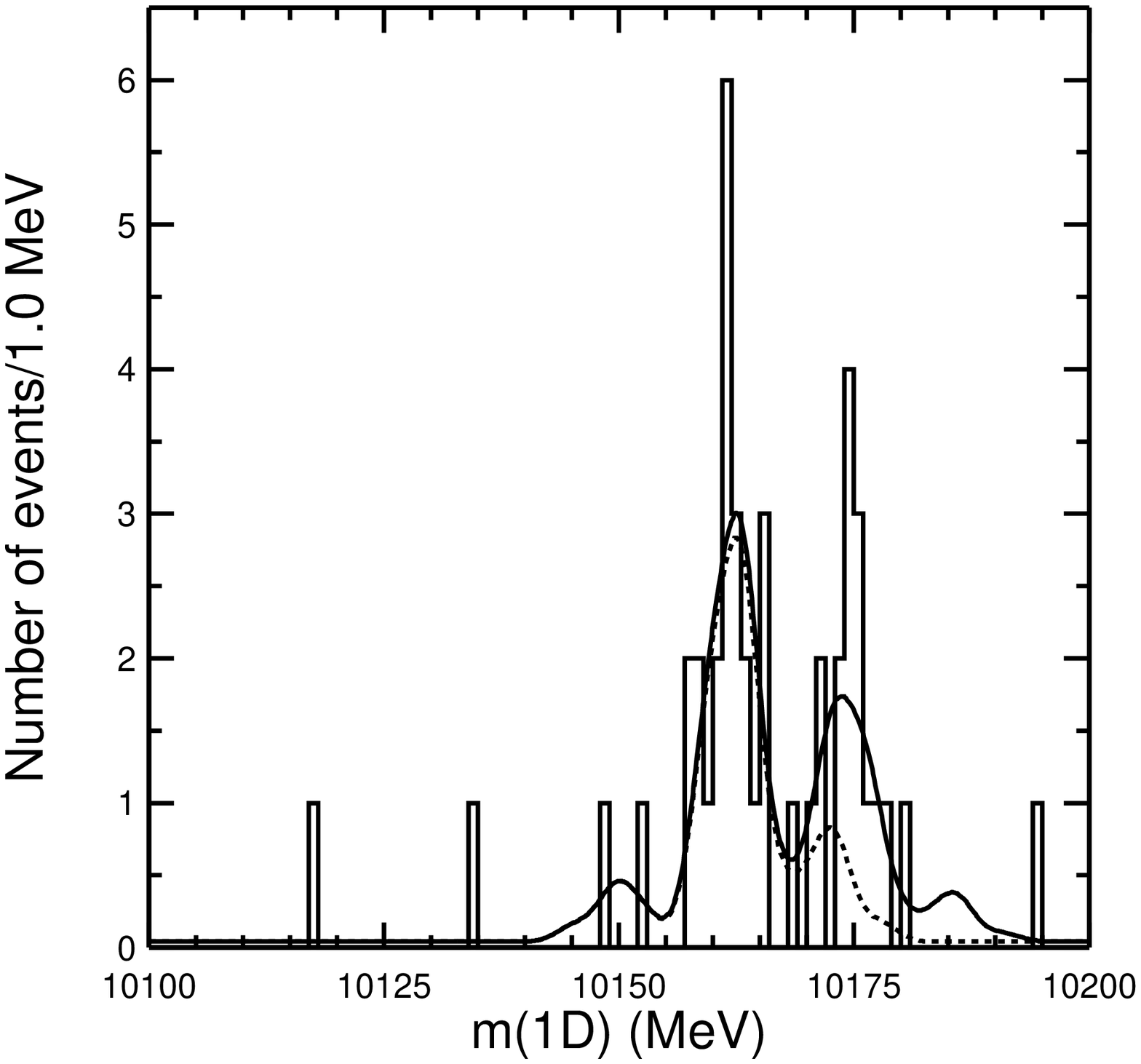,width=8cm}}
\begin{center}
\parbox{145mm}{\caption[ ]{\label{casc_12_m1dfityn.eps}\it Fit of the
Monte Carlo signal shape plus a flat background to the data. 
{\bf Left side:} 1-peak hypothesis, {\bf right side:} 2-peak hypothesis }}
\end{center}
\vspace*{-0.5cm}
\end{figure}

We fitted the data to a one-peak and two-peak hypothesis, assuming the
background to be flat. The assumption that there is no mass peak
around 10160 MeV produces low confidence level (0.04\%) and can be
ruled out at the 9.7 sigma level. The results of the fits are displayed in
Fig.~\ref{casc_12_m1dfityn.eps}. The two-peak fit gives the best
confidence level (58\%). From the change of likelihood between the
2-peak and 0-peak hypothesis we derive a significance of the peak
around 10160 MeV of 6.8 standard deviations and the significance of
the second peak is about 3 sigma. We therefore claim to see at least one state
from the $\Upsilon(1D)$ spin-triplet with sufficient significance.
The spin assignment of the state around 10160 MeV is either J=1 or J=2
since J=3 is ruled out due to our low sensitivity for that state.
Since the J=2 state is predicted to be produced with 6 times larger
rate than the J=1 state, we conclude that the J=2 state is the most
likely spin assignment with a mass of
$m(\Upsilon(1D_2)= (10161.2\:\pm\:0.7\:stat.\:\pm\: 1.0\:syst.)$ MeV
(preliminary). 
The inclusively measured product branching fraction 
$\BR(\Upsilon(3S) \to \chi_b(2P_J)) \times$\\ 
$\BR(\chi_b(2P_J) \to
\Upsilon(1D)) \times \BR(\Upsilon(1D)\to\chi_b(1P_J)) \times
\BR(\chi_b(1P_J)\to\Upsilon(1S))\times
\BR(\Upsilon(1S)\to\ell^+\ell^-)$, averaged over the $e^+e^-$ and
$\mu^+\mu^-$ modes is $(3.3\pm0.6\pm0.5)\times 10^{-5}$
(preliminary). This is in
good agreement with the theoretical predictions by Godfrey and Rosner. \cite{godfrey_rosner_y1d}
%\clearpage
\section{\label{cleoc_evt}CLEO-c}
The CLEO collaboration and CESR plan to operate in the next years 
at center-of-mass
energies in the ${\rm\tau}$/charm region\cite{yellowbook}. This will
expand the scope of our on-going charm physics program and will allow
precision tests of perturbative and lattice QCD predictions.
The results will have an impact on b-quark physics, because
heavy flavor physics, and specifically, the extraction of CKM matrix
parameters depends on our control over non-perturbative strong
interaction effects.
An appealing theory for strongly-coupled systems is lattice QCD
(LQCD). New LQCD approaches have produced a wide variety
of calculations of non-perturbative quantities with accuracies in the
1-20\% level for systems containing heavy quark(s). The techniques
needed to reduce uncertainties to 1-2\% exist, but higher
precision requires cross checks for the theory
predictions. The probably most important verification of LQCD
predictions are charm data that will be collected with CLEO-c. CLEO
has the potential to verify LQCD predictions at the 1-2\% level. The
level of verification will greatly improve the trust of the physics
community in LQCD applications.

The list of CLEO-c physics topics
is long: charm decay constants $f_D$ and $f_{D_s}$, absolute charm
branching fractions. semi-leptonic decay form factors, direct
determination of $\CKM{cd}$ and $\CKM{cs}$ with 1-2\% accuracy, spectroscopy of
charmonium states, searches for QCD exotics like hybrids and glueballs, R
measurements, rare D decays, D mixing, $\tau$
decays. \cite{shipsey_ssi,shipsey_cleoc,yellowbook}

\begin{figure}[htb]
\centerline{\epsfig{file=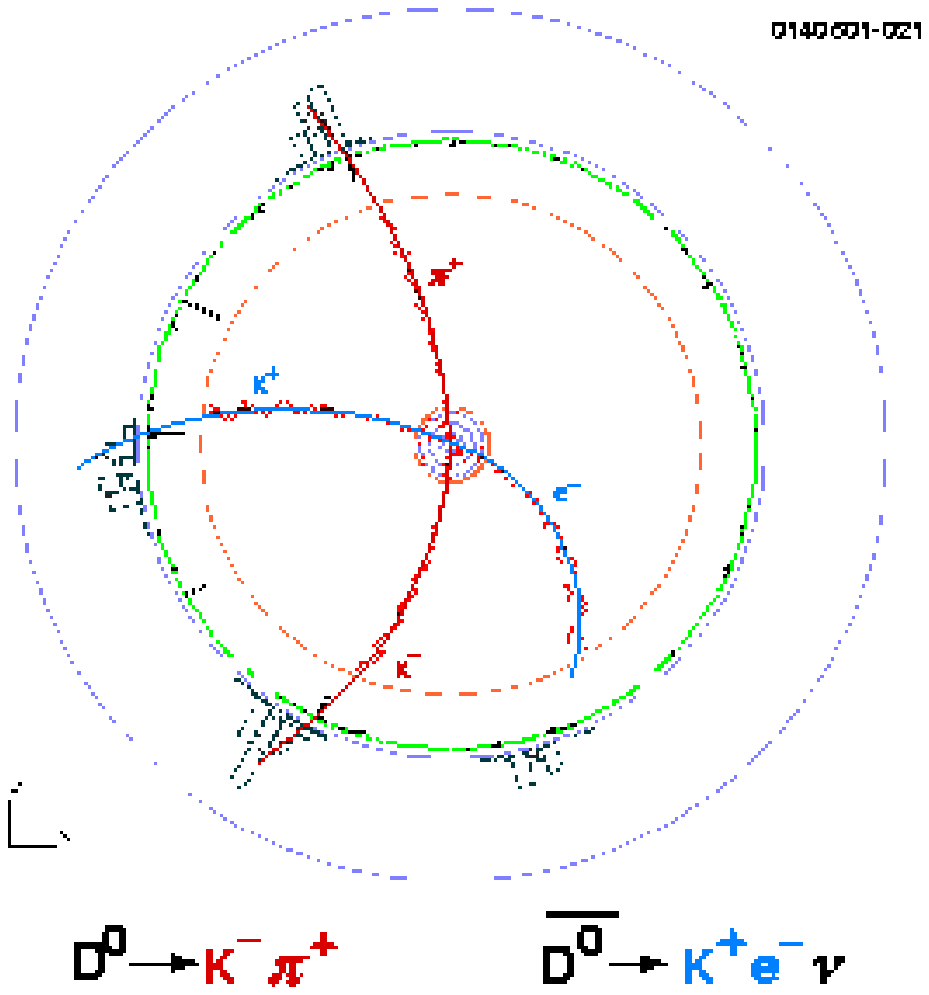,width=8cm}\epsfig{file=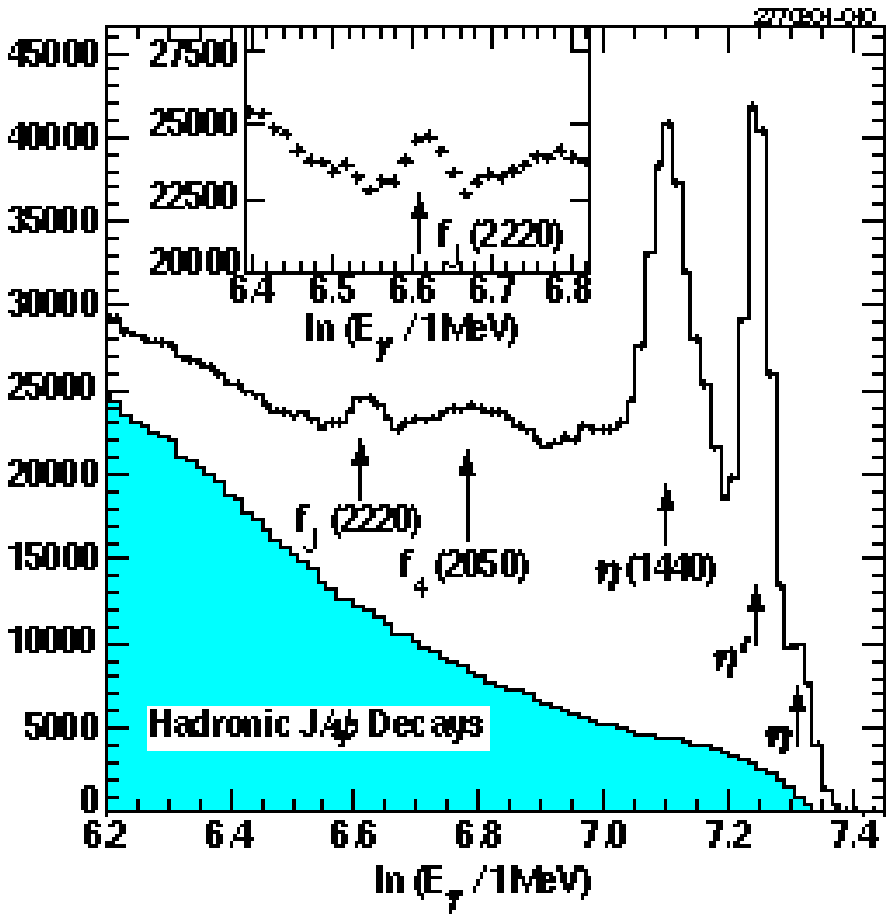,width=8cm}}
\begin{center}
\parbox{145mm}{\caption[ ]{\label{cleo-c_evtdisplay.eps}\it Left side:
CLEO-c event display of a simulated $\Psi(3770)$ decay.
Right side: Photon energy spectrum in radiate $\JPsi$ decays as a
function of $\ln(E_\gamma/1\: MeV)$. 
The background from hadronic $\JPsi$ decays is included
as the shaded area.}}
\end{center}
\vspace*{-0.5cm}
\end{figure}

These physics topics require an $e^+e^-$ collider operating on the
charmonium states $\Jpsi$ $\Psi'$ and $\Psi(3770)$. The $\Psi(3770)$
is the first $c\bar{c}$ resonance above $D\bar{D}$ threshold. The
final state is rather limited since the resonance is below threshold
for $D\bar{D}\pi$ production. The
operation of CLEO-c at this resonance would be analogous to the long and
successful running of CLEO on the \yfours.
Advantages of running there are the excellent signal to
background and the probability to tag D-mesons. The tagging is illustrated in 
Fig.~\ref{cleo-c_evtdisplay.eps}. The
flavor of the decay $D^0\to K^-\pi^+$ determines (tags) the flavor of the
recoiling D meson. Energy-momentum conservation determines the
4-momentum of the recoiling state. 

D-meson tagging makes precise 
absolute branching fraction measurements possible in addition to 
un-precedented neutrino reconstruction that is crucial for 
extracting the $D^+$ form
factor in the leptonic decay $D^+ \to \mu^+ \nu_\mu$.
Another advantage of the $\Psi(3770)$ running is the quantum coherence of the
$D\bar{D}$ system which aids $D$ mixing and CP violation studies. 

\paragraph{QCD studies with CLEO-c}
Table \ref{cleoc.tab} shows a summary of the data set size for
CLEO-c (projected) and for BES. The CLEO-c data sets will be over an
order of magnitude larger. In addition, the CLEO detector is more
modern and thus
superior to the BES II detector. This will allow us to improve on many
BES measurements due to better control over systematics. 
This can be demonstrated in two examples.\\
(1) Our hermetic detector with very good track reconstruction efficiency 
will allow us to perform exact Measurements of R, the ratio
of the ISR-corrected hadronic cross section to the first-order QED
cross section. The average uncertainty on each energy point of 
7\% (BES) can be improved with CLEO-c to $\tilde 2$\% 
in the range $\sqrt{s}=3-5$ GeV. Electroweak precision fits will
benefit from the improved R result.\cite{jegerlehner}\\
(2) The energy resolution of our CsI calorimeter is up to 20 times
better than BES-II, for example 2\% at $E_\gamma=700$ MeV. This makes
measurements of the inclusive photon spectrum in radiative \JPsi and
$\Psi'$ decays possible.  Radiative \Jpsi\ decays are an excellent
search ground for glue-rich QCD exotics. 
A CLEO-c photon spectrum from $\JPsi\to\gamma+X$ is shown 
in Fig.~\ref{cleo-c_evtdisplay.eps}, right side. The spectrum is based
on $7\times 10^7$ simulated \Jpsi\ decays. 

Narrow resonances with branching
fractions of order $10^{-4}$ can easily be identified in radiative
decays. E.g. the
peak\footnote{The $f_J(2220)$ is simulated with $\BR=8\times 10^{-4}$,
$M=2230 MeV$ and $\Gamma=23 MeV$} from the $f_J(2220)$ is clearly
visible in Fig.~\ref{cleo-c_evtdisplay.eps}.
Combined with
exclusive radiative $\Psi^{(')}$ decays, absolute branching fractions 
of narrow QCD exotics can be measured. These measurements in addition to a
full partial wave analysis of exclusive final states will elucidate
the nature of QCD exotics in the mass region below 3 GeV.
Our search for glue-rich QCD exotics 
will be complemented by a search for similar final state in
radiative Upsilon decays and an anti-search in two-photon events.
\begin{table}[htb]
\begin{center}
\begin{tabular}{r|r|r}
\hline\hline
Resonance & CLEO-c & BES-II \\ \hline \hline
$\Jpsi$ & $10^9$ & $6\times 10^7$     \\
$\Psi'$ & $10^8$ & $4\times 10^6$   \\
$\Psi(3770)$ & $3\times 10^7$ $D\bar{D}$ & -- \\
$E_{CM}=4140$ MeV & $1.5\times 10^6$ $D_s\bar{D}_s$ & $4\times 10^{5}$  \\
\hline \hline
\end{tabular}
\end{center}
\parbox{145mm}{\caption[ ]{\label{cleoc.tab} Comparison of projected
CLEO-c data samples with BES-II.}}
\end{table}

\section{Summary and Conclusions}
Since the whole text is a summary of recent CLEO results, 
another summary is not
in order. Many more interesting CLEO results on B physics will
come out in the near future. First CLEO-III results from the \yfours\
can be expected soon. CLEO has successfully finished operation on the
$\Upsilon(1S-3S)$ resonances and is soon exploring the $\tau$-charm
region. First, exploratory runs at lower energies yielded encouraging
results. High luminosity runs can be expected as soon as the CESR
accelerator upgrade in finished in 2003\cite{yellowbook}.
\paragraph{Acknowledgments}
I would like to thank my CLEO colleagues for giving me the opportunity
to present our results. I would also like to gratefully acknowledge
the  
effort of the CESR staff in providing us with excellent luminosity and
running conditions. I finally thank the organizers for
a wonderful conference experience.
%\begin{equation}
%t_{lm} = \sqrt{4\pi} \sum_{events} Y_l^m(\Omega) .  \label{eqn1}
%\end{equation}

\end{document}